\documentclass[prd,aps,twocolumn,a4paper,showkeys,nofootinbib]{revtex4-1}

\usepackage{graphicx,psfrag}
\usepackage{mathrsfs}
\usepackage{amsmath,amsfonts,amssymb}
\usepackage{multirow}
\usepackage{comment}

\usepackage{ulem}
\usepackage{hyperref}
\usepackage{enumitem}

\newcommand{\be}{\begin{equation}}
\newcommand{\ee}{\end{equation}}
\newcommand{\bea}{\begin{eqnarray}}
\newcommand{\eea}{\end{eqnarray}}
\newcommand{\bel}{\begin{align}}
\newcommand{\eel}{\end{align}}

\def\non{\nonumber}

\def\i{{\rm i}}

\def\gccm{{\rm g\,cm^{-3}}}
\def\Msun{{\rm M_{\odot}}}
\def\GMc2{{\rm G M_{\odot} c^{-2}}}

\def\w{{\hat{\omega}}}
\def\f{{\hat{f}}}
\def\M{\mathcal{M}}
\def\B{\mathcal{B}}
\def\C{\mathcal{C}}

\def\O{\mathcal{O}}
\def\MM{\bar{F}}

\def\Mo{{\rm M_{\odot}}}
\def\kt2{\kappa^\text{T}_2}
\def\Mmax{M_\text{max}^\text{TOV}}
\def\Rmax{R_\text{max}^\text{TOV}}
\def\Rmax{R_\text{max}^\text{TOV}}
\def\Mllmax{M_\text{LB}^\text{TOV}}

\usepackage{pifont} 

\newcommand{\nrpm}{\texttt{NRPM}}

\usepackage{color}
\definecolor{cyan}{rgb}{0,0.9,0.9}
\definecolor{orange}{rgb}{0.9,0.5,0}
\definecolor{magenta}{rgb}{1,0,1}
\definecolor{purple}{rgb}{0.8,0.4,0.8}
\definecolor{gray}{rgb}{0.8242,0.8242,0.8242}
\newcommand{\todo}[1]{\textcolor{red}{$\blacksquare$ TODO: #1}}

\newcommand{\bs}[1]{{\textcolor{orange}{\texttt{SB: #1}} }}

\begin{document}


\title{kiloHertz gravitational waves from binary neutron star
   remnants:\\
   time-domain model and constraints on extreme matter}

\author{Matteo \surname{Breschi}$^{1}$}
\author{Sebastiano \surname{Bernuzzi}$^{1}$}
\author{Francesco \surname{Zappa}$^{1}$}
\author{\\Michalis \surname{Agathos}$^{1}$}
\author{Albino \surname{Perego}$^{2,3}$}
\author{David \surname{Radice}$^{4,5,6,7}$}
\author{Alessandro \surname{Nagar}$^{8,9,10}$}

\affiliation{${}^1$Theoretisch-Physikalisches Institut, Friedrich-Schiller-Universit{\"a}t Jena, 07743, Jena, Germany}
\affiliation{${}^2$Dipartimento di Fisica, Universit\'a di Trento, Via Sommarive 14, 38123 Trento, Italy}
\affiliation{${}^3$Istituto Nazionale di Fisica Nucleare, Sezione di Milano-Bicocca, Piazza della Scienza 20100, Milano, Italy}
\affiliation{${}^4$Department of Physics, The Pennsylvania State University, University Park, PA 16802, USA}
\affiliation{${}^5$Department of Astronomy \& Astrophysics, The Pennsylvania State University, University Park, PA 16802, USA}
\affiliation{${}^6$Institute for Advanced Study, 1 Einstein Drive, Princeton, NJ 08540, USA}
\affiliation{${}^7$Department of Astrophysical Sciences, Princeton University, 4 Ivy Lane, Princeton, NJ 08544, USA}
\affiliation{${}^8$Centro Fermi - Museo Storico della Fisica e Centro Studi e Ricerche Enrico Fermi, Roma, Italy}
\affiliation{${}^9$Istituto Nazionale di Fisica Nucleare, Sezione di Torino, Via P.Giuria 1, 10125 Torino, Italy}
\affiliation{${}^{10}$Institut des Hautes Etudes Scientifiques, 91440 Bures-sur-Yvette, France}

\date{\today}

\begin{abstract}
  The remnant star of a neutron star merger is an anticipated loud
  source of kiloHertz gravitational waves that conveys unique
  information on the equation of state of hot matter at extreme
  densities. Observations of such signals are hampered by the
  photon shot noise of ground-based interferometers 
  and pose a challenge for gravitational-wave astronomy.
  We develop an analytical time-domain  
  waveform model for postmerger signals
  informed by numerical relativity simulations.
  The model completes effective-one-body waveforms for quasi-circular 
  nonspinning binaries in the kiloHertz regime.
  We show that a template-based analysis 
  can detect postmerger signals with a minimal signal-to-noise ratios (SNR)
  of 8.5, corresponding to GW170817-like events for third-generation interferometers. 
  Using Bayesian model selection and the complete inspiral-merger-postmerger waveform model it is
  possible to infer whether the merger outcome is a prompt collapse to a black hole
  or a remnant star. 
  In the latter case, the radius of
  the maximum mass (most compact) nonrotating neutron star can be
  determined to kilometer precision.       
  We demonstrate the feasibility of inferring the stiffness of the
  equation of state at extreme densities using the
  quasiuniversal relations deduced from numerical-relativity
  simulations. 
\end{abstract}


\maketitle




\section{Introduction} 
 
\begin{figure}[t]
  \centering 
    \includegraphics[width=0.49\textwidth]{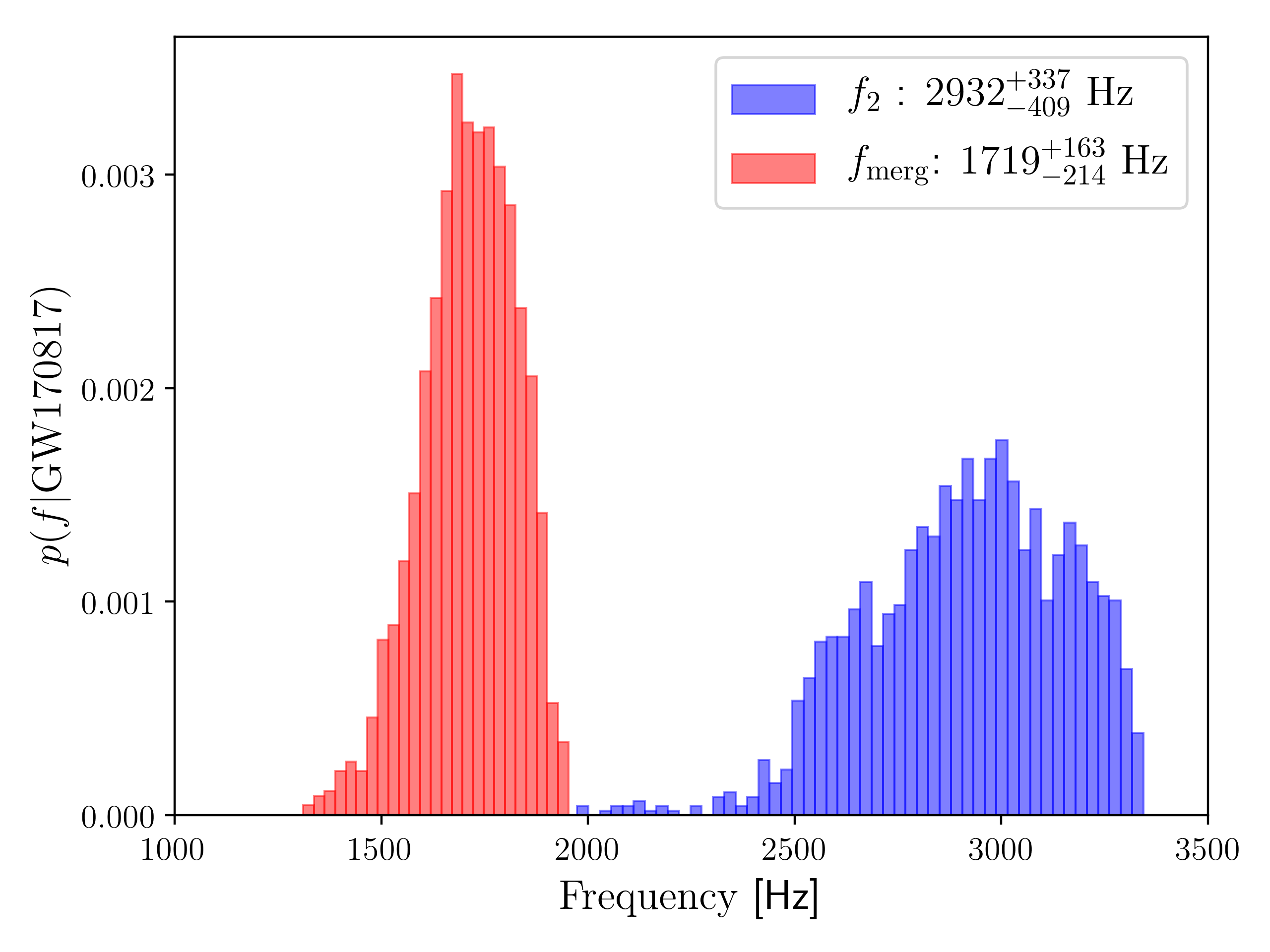}
    \caption{Gravitational-wave merger $f_\text{mrg}$ and postmerger
      peak $f_2$ frequency
      for GW170817. The distributions are estimated from the
      LIGO-Virgo posteriors distributions~\cite{LIGOScientific:2018mvr} for the $\tilde{\Lambda}$
      parameters using (i) the quasiuniversal relation proposed in 
      \cite{Bernuzzi:2014kca} for the merger frequency;
      (ii) the relation proposed in \cite{Bernuzzi:2015rla} and further refined in this work
      for the postmerger peak frequency.
      The distribution of $f_2$ is cut at $\kt2<70$ to exclude binaries that undergo
      prompt collapse at merger.}
    \label{fig:GW170817-mrg-f2}
\end{figure}

The gravitational-wave (GW) signal GW170817 is compatible with 
the inspiral of a binary neutron star (BNS) 
of chirp mass $\mathcal{M} \sim 1.186(1)\Msun$, mass ratio $q \sim [1,1.34]$ and tidal
deformability parameter distributed around $\tilde{\Lambda}\sim300$
and smaller than ${\sim}800$
\cite{TheLIGOScientific:2017qsa,Abbott:2018wiz,LIGOScientific:2018mvr}.
The merger frequency of a BNS GW can be accurately predicted using
numerical relativity (NR) results~\cite{Bernuzzi:2014kca}. From the
probability distribution of $\tilde{\Lambda}$ measured for GW170817 one finds
the merger frequency falls in the broad range $f_\text{mrg}\sim (1.2,2)$~kHz, 
Fig.~\ref{fig:GW170817-mrg-f2}. The sensitivity of the detectors in
August 2017 was
insufficient to clearly identify a signal at frequencies
$f\gtrsim f_\text{mrg}$ \cite{Abbott:2017dke,Abbott:2018hgk}.  
Indeed, LIGO-Virgo searches for short (${\lesssim}1$~s), intermediate (${\lesssim}500$~s)
and long (days) postmerger transients from a neutron star (NS) remnant
resulted in upper limits of more than one order of magnitude larger than
those predicted by basic models of quasi-periodic sources~\cite{Lai:1994ke,Cutler:2002nw,Corsi:2009jt,DallOsso:2014hpa,Lasky:2015olc,Zappa:2017xba}.
Various works have suggested that for 
GW170817-like sources postmerger frequencies are accessible only by improving the design 
sensitivity of current detectors of a factor two-to-three or with
next-generation detectors~\cite{Clark:2015zxa,Abbott:2017dke,Torres-Rivas:2018svp,Martynov:2019gvu}. 

NR simulations predict that BNS mergers can form a
black hole (BH) from gravitational collapse of the merged object or a NS remnant depending on the
binary mass and the NS matter equation of state (EOS), e.g.~\cite{Shibata:1999wm,Kiuchi:2009jt,Hotokezaka:2011dh,Bauswein:2013jpa,Dietrich:2016hky,Dietrich:2016lyp}.
NS remnants can  collapse on dynamical (${\sim}\O(10)$~ms,
short-lived remnant) or longer timescales (long-lived remnant), but can also reach a
stable NS configuration.
KiloHertz GWs contain the
imprint of the merger remnant dynamics. The main signature is
a short GW transient peaking at a few characteristic frequencies, the
dominant one being associated with twice the rotation frequency of the
remnant NS at $f_2>f_\text{mrg}$~\cite{Shibata:2002jb,Stergioulas:2011gd,Bauswein:2011tp,Bauswein:2012ya,Hotokezaka:2013iia,Takami:2014zpa,Bernuzzi:2015rla,Radice:2016gym,Lehner:2016lxy,Dietrich:2016hky,Dietrich:2016lyp}. The transient
is more luminous for short-lived remnant than for long-lived; an
absolute upper limit to the energy per unit mass is  
${\lesssim} 0.126 (\frac{M}{2.8\Msun})~\Msun {\rm c}^2$, where $M$ is the binary
mass \cite{Zappa:2017xba}.
Long postmerger transients are also possible for NS remnants developing
nonaxisymmetric instabilities and/or
magnetars, but they are expected to be less luminous than the GWs on dynamical
timescales, e.g. \cite{Lai:1994ke,Cutler:2002nw,Corsi:2009jt,DallOsso:2014hpa,Lasky:2015olc}.
Recent analysis of GW170817 based on
premerger GWs combined with the pulsar constraints on the maximum
mass largely disfavor
prompt collapse to BH~\cite{Agathos:2019sah}. 
Using the NR relation between the frequency $f_2$ and the 
tidal deformability derived in~\cite{Bernuzzi:2015rla} and the LIGO-Virgo posteriors
for GW170817, one finds that a tentative wave with peak luminosity 
larger than $0.1 \times 10^{56}~{\rm erg}\cdot{\rm s}^{-1}$
could have been detected at $f_2\sim[2.5,3.2]$~kHz
(Fig.~\ref{fig:GW170817-mrg-f2}) if the instruments were more sensitive.
This is compatible with the interpretation of the electromagnetic
counterparts that suggests the formation of a short-lived NS
remnant~\cite{Margalit:2017dij,Bauswein:2017vtn,Shibata:2017xdx,Radice:2017lry,Rezzolla:2017aly},
although other scenarios are possible
\cite{Yu:2017syg,Ai:2018jtv,Li:2018hzy,Lazzati:2016yxl,Bromberg:2017crh}.

The data analysis of (short duration) postmerger signals can be performed with either morphology independent
approaches \cite{Chatziioannou:2017ixj,Torres-Rivas:2018svp} or using matched filtering
techniques based on waveform templates. While matched filtering is
proven to be an optimal method in case of gaussian noise~\cite{Helstrom:1968}, its performance
for postmerger analysis 
remains unclear due to the uncertainties of postmerger templates. 
Current postmerger models comprise frequency-domain
statistical representation of NR waveforms
\cite{Clark:2015zxa,Easter:2018pqy} or simple analytical models 
\cite{Hotokezaka:2013iia,Bauswein:2015vxa,Bose:2017jvk,Tsang:2019esi}. A common
aspect of all these approaches is the 
use of NR information in terms of quasiuniversal (EOS independent)
relations for the characteristic frequencies
\cite{Bauswein:2011tp,Bauswein:2014qla,Takami:2014zpa,Bernuzzi:2015rla,Lehner:2016wjg,Rezzolla:2016nxn,Kiuchi:2019kzt}.
The relevance of these relations is twofold: on one hand they are used
for waveform modeling, on the other hand they can be used to extract
information from the analysis.

Observations of kiloHertz GWs from NS remnants 
can deliver constraints on the EOS of matter
in a regime at which nuclear interactions are still very uncertain.
For a canonical binary of mass $M=(1.4+1.4)\Mo$, tidal interactions in the
inspiral-merger part of the GW signal mostly inform about the EOS at 
about twice the nuclear saturation density $\rho_0\simeq2.3\times10^{14}~\gccm$,
corresponding to the maximal densities of the binary components
\cite{Abbott:2018exr,Agathos:2019sah}. 
However, NS remnants formed in mergers
reach densities ${\sim}3-5\rho_0$ and temperatures in excess of
${\sim}50$~MeV, e.g.~\cite{Perego:2019adq}.
The strongest constraints on the EOS at those extreme densities are currently
provided by the mass measurements of two pulsars in binary systems~\cite{Antoniadis:2013pzd,Cromartie:2019kug}.
The latter give lower bounds for the maximum mass of nonrotating
stable NS in equilibrium ($\Mmax$, hereafter simply referred as the maximum NS mass):
$\Mmax\gtrsim(2.01\pm0.04)\Mo$ (PSR J0348+0432) \cite{Antoniadis:2013pzd} and
$\Mmax\gtrsim(2.17\pm0.11)\Mo$ (PSR J0740+6620) \cite{Cromartie:2019kug}.

Additional constraints on matter at extreme densities can be inferred from the
kiloHertz GW from merger remnants by extracting NS properties via
quasiuniversal relations \cite{Bauswein:2014qla,Bernuzzi:2015rla,Steiner:2015aea}. 
Moreover, new degrees of freedom or matter phases at ${\sim}3-5\rho_0$ can impact the
remnant dynamics and leave detectable imprints on the GW. Case
studies considered matter models including hyperon production 
\cite{Sekiguchi:2011mc,Radice:2016rys} or zero-temperature models of
phase transitions to quark-deconfined matter
\cite{Bauswein:2018bma,Most:2018eaw}. The detectability of these
effects crucially depends on the densities at which the EOS softening
(or stiffening) takes place and would in principle need detailed waveform
models that are presently not available.
\\

In this paper we construct the first phase-coherent
inspiral-merger-postmerger model for the BNS GW spectrum and
demonstrate its applications to constrain the NS EOS in GW astronomy
observations. 

Section~\ref{sec:NRPM} introduces a NR postmerger model
for quasi-circular binaries called \nrpm{}, based on the quasiuniversal
relations of~\cite{Bernuzzi:2015rla} and implemented using the NR database of
the computational relativity ({\tt CoRe}) collaboration
\cite{Dietrich:2018phi}. 

Section~\ref{sec:validation} discusses performances of \nrpm{}
using a validation set of NR simulations. Section~\ref{sec:IMPM}
discusses how to complete effective-one-body 
waveforms with \nrpm{} in order to obtain a phase-coherent model of
the complete inspiral-merger-postmerger waveform, valid from the circular 
adiabatic regime to the kiloHertz regime.

Section~\ref{sec:injection} demonstrates the use of the model in
template-based Bayesian data analysis applications.
We discuss the minimal requirement for postmerger detection.  We demonstrate how to
infer prompt collapse using our complete spectrum model and Bayesian model selection. 
We show how to set constraints on the minimum NS radius from a single
event. Finally, we discuss how to infer EOS 
stiffness at the extreme densities reached in the merger remnant.  

\paragraph*{Conventions}

For waveform modeling we mostly use geometric units $c=G=1$ and
measure masses in terms of Solar masses $\Msun$. 
The waveform strain is decomposed in multipoles as 
\be
 h_+ - \i h_\times
  =D_L^{-1}\sum_{\ell=2}^\infty\sum_{m=-\ell}^{\ell} h_{\ell m}(t){}_{-2}Y_{\ell m}(\iota,\psi),
\ee
where $D_L$ is the luminosity distance and ${}_{-2}Y_{\ell m}$ are the $s=-2$
spin-weighted spherical harmonics. In this paper we shall compute the strain
from the equation above assuming only the $\ell=2$, $m=\pm 2$ modes and
symmetry across the orbital plane\footnote{We are considering here only 
nonprecessing systems.}.  The $\ell=m=2$ waveform mode is decomposed 
in amplitude $A(t)$ and phase $\phi(t)$ as
\be
h_{22}(t) = A(t)\exp{\left(- \i \phi(t)\right)} \  \ ; \ \ \omega(t) = \dot{\phi}(t) \ ,
\ee
where $\omega(t)$ also indicates the GW frequency and the dot denotes the time derivative.
The corresponding spherical harmonics are 
\be
{}_{-2}Y_{2,\pm2}(\iota,\psi)=\sqrt{\frac{5}{64\pi}}\big(1\pm\cos(\iota)\big)^2\,e^{\pm2\i\psi}\,, 
\ee
so that one obtains
\begin{align}
  &h_+ - \i h_\times\approx 
\sqrt{\frac{5}{4\pi}} \frac{A(t)}{D_L} \nonumber\\
&\times\left[
\frac{1}{2}\left(\cos ^2(\iota)+1\right) \cos(\phi(t))
- \i \cos(\iota)\sin(\phi(t))
\right] \non\ ,
\end{align}
where one sets $\psi=0$.
We work with quantities rescaled by the total binary mass, i.e.
\be
\w := M\omega = 2 \pi \f\ , \ \
\hat{t} := t/M \ , \ \
\hat{A} := A/M \ ,
\ee
and further define the moment of merger ($\hat{t}_\text{mrg}=0$) as the time of
the peak of $A(t)$ (Fig.~\ref{fig:merger}).
Note that the time $\hat{t}$ refers to the retarded time in case of the NR data.
The binary mass is indicated with $M= M_A +
M_B$, the mass ratio $q = M_A /M_B \ge 1$ and the 
symmetric mass ratio $\nu = M_A M_B / M^2$.
GW spectra and frequencies are instead discussed and shown in SI
units with distances expressed in Mpc.


\section{\nrpm{} model} 
\label{sec:NRPM}

\begin{figure}[t]
  \centering 
    \includegraphics[width=0.49\textwidth]{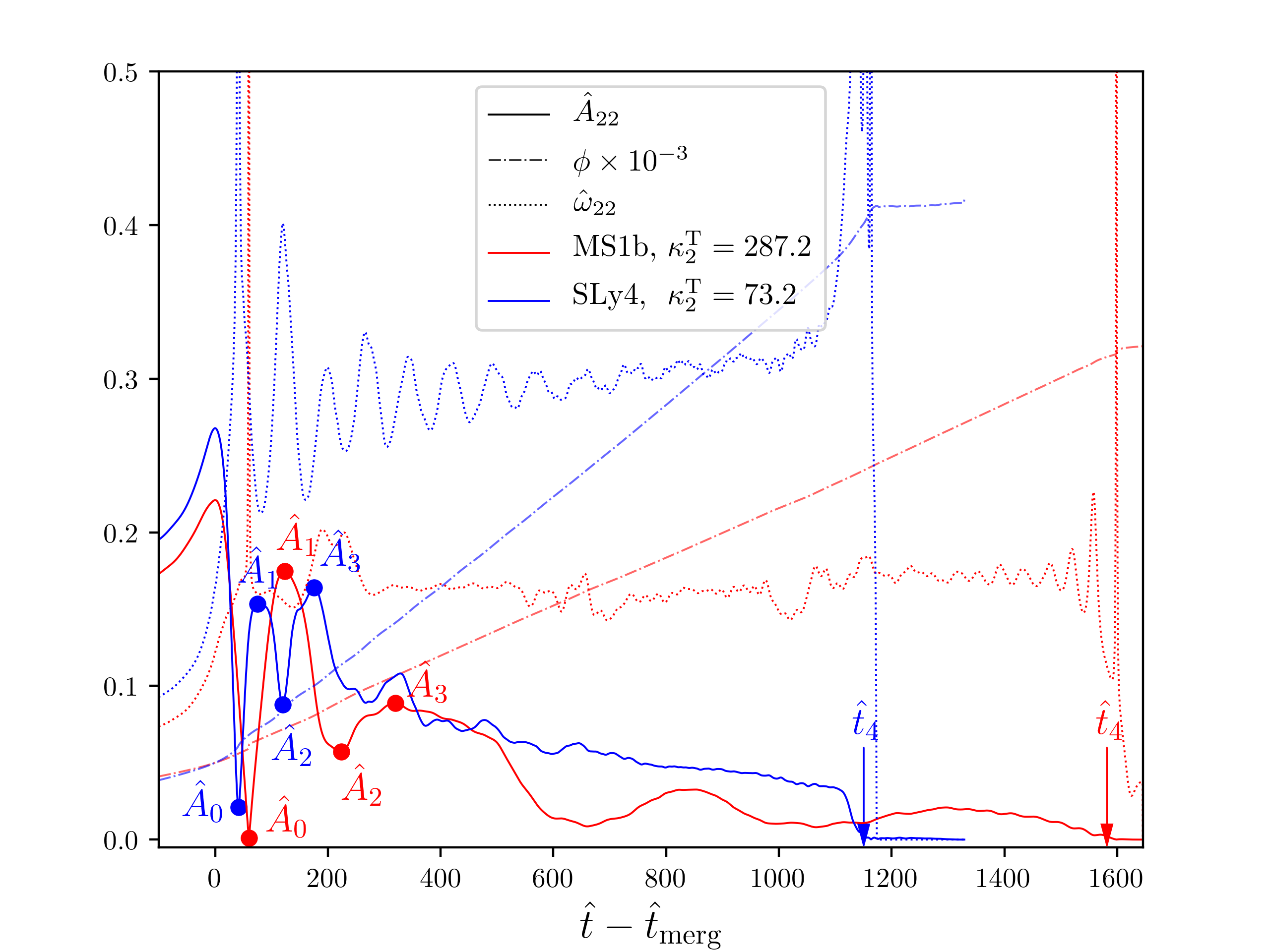}
    \caption{Merger and postmerger waveform from two very different
      BNS with mass $M=(1.35+1.35)\Mo$. The MS1b BNS is an example of
      long-lived remnant; the SLy BNS an example of short-lived
      remnant collapsing at $\hat{t} \sim 1200$ after merger time, 
      $\hat{t}=\hat{t}_\text{mrg}$.
      In both cases the postmerger waveform amplitude has
      characteristics maxima and minima $\hat{A}_i$ at times 
      $\hat{t}_i$ with $i=0,...,3$. Note the jump in the phase at
      $\hat{t}_0$, where the instantaneous frequency is not defined.}
    \label{fig:merger}
\end{figure}

Our postmerger model builds on the results of \cite{Bernuzzi:2015rla,Zappa:2017xba,Dietrich:2018uni}
that showed the postmerger frequency peak correlates with the tidal
polarizability parameter 
\be\label{eq:kappa2T}
 \kappa_2^\text{T} = \dfrac{3}{2} \left[ \Lambda_2^\text{A} \left(X_\text{A}\right)^4 X_\text{B} + \Lambda_2^\text{B} \left(X_\text{B}\right)^4 X_\text{A} \right] \ ,
\ee
%
%
%
where $\Lambda_2^i\equiv 2 k^i_2 (M_i/R_i)^{5}/3$, with $i=( A,B)$, 
are the dimensionless quadrupolar tidal polarizability parameters of the individual 
stars~\cite{Flanagan:2007ix,Damour:2009wj}, $k^i_2$ 
the dimensioless quadrupolar Love numbers~\cite{Damour:1983a,Hinderer:2007mb,Damour:2009vw,Binnington:2009bb}, 
$(M_i,R_i)$ the mass and radius and $X_\text{i}\equiv M_\text{i}/M$.
Here we derive similar relations also for other characteristic
frequencies of the spectrum and for the waveform's amplitudes and
characteristic times.
For nonspinning and slowly spinning BNS, each of
those quantities can be approximately modeled in terms of the following set of physical parameters 
\be\label{eq:theta}
\boldsymbol{\theta} = \left(\nu,M,\kt2\right) \ ,
\ee
that defines \nrpm{}'s parameter space. 
The latter choice is one of the key differences with respect to
previous time-domain models~\cite{Hotokezaka:2013iia, Bauswein:2015vxa, Bose:2017jvk}.
Other important differences are the use of the largest-to-date set of
NR simulations and the possibility of constructing a time domain
approximant that is phase coherent with inspiral-merger models (see Sec.~\ref{sec:IMPM}).

We use 148 simulations of the computational relativity (\texttt{CoRe}) 
collaboration \cite{Dietrich:2018phi}, plus 24 simulations in part 
reported in \cite{Radice:2018pdn} and in part unpublished.
The set of simulations covers the range $q \in [1 , 1.5] $ and
$\kt2 \in [73 , 458]$.

Figure~\ref{fig:merger} illustrates some of the qualitative
features common to all the merger+postmerger NR waveforms for short-
and long- lived NS remnants.
The waveform frequency at early times is approximately constant around the
$\f_2$ value. In many waveforms a further
frequency modulation is clearly present in the first 
milliseconds after merger. This feature is interpreted as the couplings 
between $\f_2$ and a radial pulsation mode $\f_0$, in analogy to what happens
with nonlinear perturbations of equilibrium NS
\cite{Dimmelmeier:2005zk,
  Passamonti:2007tm,Baiotti:2008nf,Stergioulas:2011gd}.
%
%
In the latter case, nonlinear couplings between proper modes result in new
frequencies given by $\f_{2\pm0}=\f_2\pm\f_0$. In the case of BNS mergers,
the two secondary peaks in the GW spectra can be interpreted as the
nonlinear pulsations of the remnant $\f_{2\pm0}$ \cite{Stergioulas:2011gd}.
These secondary frequency peaks in the spectrum are
well-studied, e.g.~\cite{Takami:2014tva,Bauswein:2015yca,Dietrich:2016hky,Dietrich:2016lyp}
and can be clearly seen in Fig.~\ref{fig:fullmodel}.

Although we will often refer to discrete frequencies (spectral
peaks), we stress that the GW frequency is not constant but evolves 
(chirp-like) as the remnant becomes more compact and eventually collapses 
(see SLy data in Fig.~\ref{fig:merger}). The largest GW luminosity is emitted
at early times after merger at which $\f(t)$ is approximated by a certain
combination of $\f_2,\f_{2\pm0}$ \cite{Bernuzzi:2015opx}.
The waveform's amplitude after the merger peak has typically a
minimum, a maximum and at least a second oscillation. In
Fig.~\ref{fig:merger} these extrema are labelled as
$\hat{A}_i$ and occur at times $\hat{t}_i$ with $i=0,1,2,3$ where the minima have even indices.
Note that at $\hat{t}_0$ the GW phase has a jump and the instantaneous
frequency is not defined; this corresponds to a moment in which the
remnant has a strongly suppressed quadrupolar deformation. At timescales ${\sim}10-20$~ms
corresponding to $\hat{t} \sim 1000-2000$ ($M\sim2.7\Msun$) the
remnant has either collapsed (short-lived) or dissipated most of its 
energy via GWs. There is no significant GW emission at timescales
$\tau\gtrsim100$~ms \cite{Radice:2016gym,Ciolfi:2019fie} (see also Appendix~\ref{app:long-conv}).

In the following we describe in detail the construction of the time-domain model
and how the NR information is extracted.

\subsection{Time-domain model}
\label{sbsec:TDmodel}

\subsubsection{Frequency and Phase}

We assume the GW frequency is composed of the three
main characteristic frequencies $\f_{2-0}<\f_2<\f_{2+0}$ and construct a
$\C^1$ model for $\w(t)$ as follows. The frequency
model starts at $\hat{t}=\hat{t}_\text{mrg}=0$ with the value of the
merger frequency $\w_\text{mrg}$ and its derivative
$\dot{\w}_\text{mrg}$ taken either from NR fits or from an
inspiral-merger time-domain approximant (see Sec.~\ref{sec:IMPM}).
We impose
\begin{subequations}
\begin{align}
\w(\hat{t}_\text{mrg}) &= \w_\text{mrg} \\
\w(\hat{t}_0\leq\hat{t}\leq\hat{t}_1) &= \w_{2-0} \\
\w(\hat{t}_2) &= \w_{2+0} \\
\w(\hat{t}\geq\hat{t}_3) &= \w_{2} \ , 
\end{align}
\end{subequations}
and use a cubic interpolant to join $\w_\text{mrg}$ to $\w_{2-0}$
in the interval $(\hat{t}_\text{mrg},\hat{t}_0)$ fixing the values of
the function and of the first derivatives at the interval's
extrema. The derivative at $\hat{t}_0$ is taken as 
$\dot{\w}(\hat{t}=\hat{t}_0)=0$. 
The frequency oscillation in the intervals $(\hat{t}_1,\hat{t}_2)$ and 
$(\hat{t}_2,\hat{t}_3)$ is modeled with a sine function in such a way that 
$\w_{2+0}$ is a maximum and preserving the continuity and the
differentiability of $\w(t)$.
Note the model can be reduced to a single-frequency one by simply
joining $\w_\text{mrg}$ to $\w_2$ at $\hat{t}_3$ and 
omitting $\w_{2\pm0}$.
The phase of the waveform is finally given by integrating the
frequency model, 
\be
 \phi (\hat{t}) = \int_{0}^{\hat{t}} \w(\hat{t}') \mathrm{d}\hat{t}'+\phi_0 \ ,
\ee
where $\phi_0$ is either arbitrary chosen or fixed by requiring continuity
with an inspiral-merger phase. 
 
\subsubsection{Amplitude}

We assume the postmerger amplitude has two minima, $\hat{A}_i$ with $i=0,2$, 
and two maxima, $\hat{A}_i$ with $i=1,3$, and that it decays
exponentially after the second maximum. A $\C^1$ model for
$\hat{A}(t)$ is constructed assuming
\begin{subequations}
\begin{align}
\hat{A}(\hat{t}_\text{mrg}) &= \hat{A}_\text{mrg} \\
\hat{A}(\hat{t}_i) &= \hat{A}_{i} \\
\hat{A}(\hat{t}\geq\hat{t}_3+5) &= \hat{A}_3 \exp{\left[- \alpha \left(\hat{t}-\hat{t}_3\right)\right]}
\ , 
\end{align}
\end{subequations}
and using sine waves to connect maxima and minima. We define
fractional amplitudes $\beta_i=\hat{A}_i/\hat{A}_\text{mrg}$ with 
$i=0,1,2,3$ of the extrema with respect to the merger amplitude. The
damping term $\alpha$ is set as the time scale at which the waveform
amplitude is $1/100$ of the merger value, i.e. when $\hat{A}$ falls
below the threshold 
\be
\beta_4 = 10^{-2} \ .
\ee
Indicating $\hat{t}_4$ such time, one obtains
\be
\label{eq:alpha}
\alpha = \frac{\ln(100\, \beta_3)}{\hat{t}_4 - \hat{t}_3} \ .
\ee
The timescale $1/\alpha$ is identified from
simulations and has range ${\sim}(3,70)$~ms for BNS masses distributed
$M \sim (2.5,3)\Msun$, if no collapse to a BH
happens before \cite{Bernuzzi:2015opx} (see also Sec.~\ref{sbsec:PromptCollapse} for discussion on BH collapse).

\subsection{NR information} 
\label{sbsec:NRinfo}

\begin{table*}[t]
  \centering    
  \caption{\nrpm{} model parameters and their ranges, 
    coefficients of NR fits with rational functions 
    ($F_0$,$n_1$,$n_2$,$d_1$,$d_2$) or with linear functions ($p_0,p_1$), and
    fits' $\chi^2$. }
    \resizebox{\textwidth}{!}{
  \begin{tabular}{ccccccccccccc}        
    \hline\hline
    Parameter & Description & Range & NR fit model & $c$ & $F_0$ & $n_1$ & $n_2$& $d_1$ & $d_2$ & $p_0$ & $p_1$ & $\chi^2$\\
    \hline
    $\f_\text{mrg}$ & Merger frequency & $[ 0.013872, 0.027953]$ & Rational & $3199.8$ & $0.033184$ & $0.0013067$ & $0.00$ & $0.0050064$ & $0.00$ & - & - & $1.539\times10^{-5}$  \\
    $\f_2$ & PM peak frequency & [0.021789, 0.048804] & Rational &-52.655 & 7.6356 & 0.066645 & 4.0146 $\times$ 10$^{-5}$ & 10.949 & 0.040276 &  & - &$9.702\times10^{-5}$\\
    $\f_{2-0}$ & PM secondary frequency & [0.013756, 0.037838] & Rational & 5767.6 & 0.052182 & 0.002843 & 0.00 & 0.012868 & 0.00 & - & - &$1.033 \times 10^{-4}$\\
    $\f_{2+0}$ & PM secondary frequency & [0.029628, 0.071988] & Rational & 1875.5 & 4.5722 & 0.060385 & 1.0661 $\times$ 10$^{-4}$& 4.1506 & 0.027552 &- & - & $5.213\times 10^{-4}$ \\
    \hline
    $\hat{A}_\text{mrg}$ & Merger amplitude & [0.17296, 0.27331] &  Rational & 5215.0 & 0.34910 & 0.019272 & -4.3729 $\times$ 10$^{-6}$ & 0.028266 & 9.3643 $\times$ 10$^{-6}$ & - & - &$1.421\times 10^{-4}$ \\
    $\hat{A}_0$ & 1st mininum of PM amplitude & [0.0023760, 0.049993] &  Linear & -6735.8 & - & - & - & - & - & 0.032454 & -6.8029 $\times$ 10$^{-5}$ & $3.877\times 10^{-3}$ \\
    $\hat{A}_1$ & 1st maxinum of PM amplitude & [0.059723, 0.21650] &  Linear & 58542.0 & - & - & - & - & - & 0.17657 & -3.7794 $\times$ 10$^{-5}$ & $1.308\times 10^{-3}$ \\
    $\hat{A}_2$ & 2nd mininum of PM amplitude & [0.016075, 0.15814]&  Linear & -623.09 & - & - & - & - & - & 0.11601 & -1.7376 $\times$ 10$^{-4}$ & $4.700\times 10^{-3}$  \\
   $\hat{A}_3$ & 2nd maxinum of PM amplitude & [0.049711, 0.19158] &  Linear & 4486.2 & - & - & - & - & - & 0.15894 & -1.7317 $\times$ 10$^{-4}$ &  $4.177\times 10^{-3}$ \\
    \hline
    $\hat{t}_\text{mrg}$ & Merger time & 0 & - & - & - & - & - & - & - & - & - & - \\
    $\hat{t}_0$ & Time of $\hat{A}_0$ &$[39.488, 77.146]$ & Linear & 241.88 & - & - & - & - & - & 37.181 & 0.086789 & 0.1509 \\
    $\hat{t}_1$ & Time of $\hat{A}_1$ &$[56.489, 162.76]$ &Linear & -4899.3 & - & - & - & - & - & 83.045 & 0.16377 &  2.124 \\
    $\hat{t}_2$ & Time of $\hat{A}_2$ &$[71.284, 416.15]$ & Linear & -6027.2 & - & - & - & - & - & 121.34 & 0.3163 & 18.17 \\
    $\hat{t}_3$ & Time of $\hat{A}_3$ &$[87.423, 506.15]$ & Linear & -6312.6 & - & - & - & - & - &157.29 & 0.48347 &18.28 \\
    $\hat{t}_4$ & Time of $\hat{A}=\hat A_{\rm mrg}\times 10^{-2}$ & [264.14, 5011.6] & Linear & 8573.6 & - & - & - & - & - & 1375.0 & 1.8460  & 413.3\\
    \hline\hline
  \end{tabular}}
 \label{tab:fitcoefs}
\end{table*}

\begin{figure*}[t]
  \centering 
    \includegraphics[width=\textwidth]{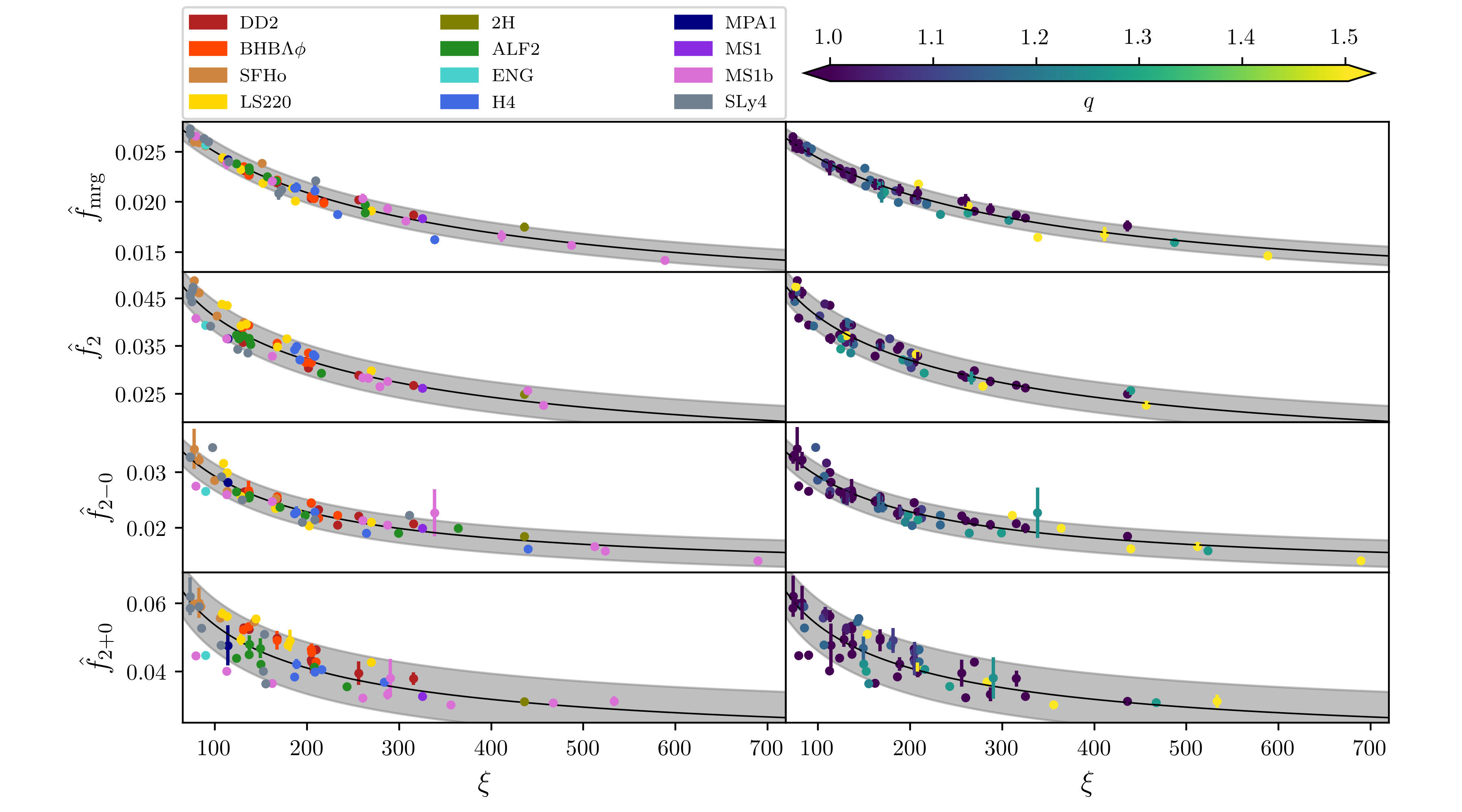} 
    \caption{Characteristic frequencies information from NR
      simulations. Markers represent the frequencies 
      extracted from the NR data and the uncertainties are estimated using 
      simulations at different resolutions;
      the black lines are the fits and the grey bands are the 90\% credible regions.
      Left and right panels show the same data: the colors on the left panel
      correspond to the EOS variation, on the right panel to the mass ratio.}
    \label{fig:freqfit}
\end{figure*}

The model's parameters are summarized in Tab.~\ref{tab:fitcoefs}; 
their values are fixed by constructing interpolating formulas of NR
data on the space of parameters $\boldsymbol{\theta}$.

\subsubsection{Frequencies, amplitudes and times}
\label{sbsec:NRinfo:fat}

The frequency information is extracted from the
spectra by identifying the three dominant peak frequencies.
Amplitudes $\hat{A}_i$ and the related times $\hat{t}_i$ are
extracted from the waveforms (Fig.~\ref{fig:merger}). Specifically,
we construct fit models using the variable \cite{Zappa:MSc} (see also
Appendix~\ref{app:quni-rel}) 
\be\label{eq:xi}
\xi = \kappa_2^T + c (1-4 \nu) \ ,
\ee
where the constant $c$ is also a fitting parameter.
The frequency and amplitude at merger
$\hat{A}_\text{mrg}$, and the peak frequencies are well described by 
rational functions in the form 
\be
F_\text{Rational}(\kt2,q) = F_0 \frac{1 + n_1 \xi + n_2 \xi^2}{1 + d_1 \xi+ d_2 \xi^2} \ ,
\ee 
where $(F_0,n_1,n_2,d_1,d_2)$ are the fitting parameters.
The amplitudes $\hat{A}_i$ for $i = 0,1,2,3$ and the times
$\hat{t}_i$ are instead fit by linear polynomials in $\xi$
\be
F_\text{Linear}(\kt2,q) = p_0 + p_1\xi \ ,
\ee
where $(p_0,p_1)$ are fitting parameters.
The results of the fits are shown in Tab.~\ref{tab:fitcoefs}. 

As an example, the peak frequency fits are shown in
Fig.~\ref{fig:freqfit}.
The uncertainty of the NR data computed from simulations at multiple
grid-resolutions is shown in the plot as bars, if available.
Note the $\hat{f}_2$ peaks determination is affected by a further error of
${\sim}{2-8\%}$ due to the discrete Fourier transform; larger
errors affect the $\f_{2\pm0}$ determination.
The $\chi^2$ coefficients for the frequencies fit are typically
${\sim}10^{-4}$ (note the merger frequency has $\chi^2\sim10^{-5}$),
but some outliers are visible from the plots at small $\xi$,
or equivalently small $\kt2$ (since these points correspond to $q\sim 1$).
We note that most of these data points correspond to
low-resolution simulations for which error bars either cannot be
computed (one resolution available) or are unreliable (two low
resolutions available). For example, the ENG simulation at $\kt2\sim80$
is a high-mass $M=(1.7+1.7)\Msun$ BNS simulated at a maximal grid resolution
of $h\approx0.365$~km that does not guarantee convergence even for the 
inspiral-merger 
(cf.~\cite{Bernuzzi:2013rza,Radice:2013xpa,Bernuzzi:2016pie} and Appendix~\ref{app:long-conv}). 
The frequency $\f_{2+0}$ model is the most uncertain for the available data. 

Table~\ref{tab:fitcoefs} (see also Appendix~\ref{app:quni-rel}) shows
that, while postmerger amplitude fits are well captured by the model
($\chi^2\sim10^{-3}$), the postmerger times are more uncertain
($\chi^2>1$) with the uncertainty growing for larger times. This is expected
since the quantities at later times are less correlated with
pre-merger parameters and NR data are themselves more uncertain the
longer the simulation is. While uncertainties on ``late-time''
quantities do not affect significantly the time-domain waveform (see discussion in
Sec.~\ref{sec:validation}), they can affect the Bayesian parameter
estimation (Sec.~\ref{sec:injection}). Notably, the damping parameter
$\alpha$ is degenerate with part of the waveform amplitude in Fourier space, and
therefore fit biases can affect the estimation of the luminosity distance.

\subsubsection{Prompt collapse}
\label{sbsec:PromptCollapse}

NR simulations indicate that a NS binary merger will
be followed by a prompt collapse to a BH, if the total
gravitational mass $M$ of the binary exceeds a threshold mass. The
latter can be roughly estimated as \cite{Hotokezaka:2011dh,Bauswein:2013jpa} 
\be\label{eq:Mthr}
M_\text{thr} = k_\text{thr} \Mmax \ .
\ee
where $\Mmax$ is the gravitational mass of the
heaviest stable nonrotating NS. Both $\Mmax$ and
$k_\text{thr}$ depend, in general, on the EOS, mass ratio, and spins.
For a sample of hadronic EOS and equal-mass nonspinning binaries,
the threshold parameter in Eq.~\eqref{eq:Mthr} is found in the range
$1.3 \lesssim k_\text{thr}\lesssim 1.7$ 
\cite{Hotokezaka:2011dh,Bauswein:2013jpa,Agathos:2019sah}.
Moreover, $k_\text{thr}$ shows an approximately EOS-independent linear
behaviour in the compactness $C$ of a reference nonrotating
NS at equilibrium, see \cite{Agathos:2019sah} for a recent collection of
literature data, fit recalibration and discussion.
Despite several NR efforts, it remains challenging to construct a
EOS-independent (universal) relation for $M_\text{thr}$ that is accurate 
and robust across the entire parameter space. A data analysis approach
based on Eq.~\eqref{eq:Mthr}, \nrpm{} and EOS inference is outlined in
Appendix~\ref{app:EOS-inference-pc}.

We follow here an alternative route.
By analyzing the NR data of the {\tt CoRe} collaboration, 
we have found that all the 30 prompt collapse mergers are captured by the condition $\kt2 <80$,
see also Ref.~\cite{Zappa:2017xba}.
By further combining the estimate with Eq.~\eqref{eq:Mthr} for a
sample of nonrotating NS model with 13 EOS, leads to the
following criterion for prompt collapse \cite{Zappa:2017xba}
\be
\label{eq:k_thr}
\kt2 < \kappa^{\rm T}_\text{thr} = 80\pm40 \ .
\ee
We adopt the above criterion in \nrpm{}.
In the context of a Bayesian analysis, the threshold value can be either
prescribed or included in the set of intrinsic parameters.

This assumption is a simplification as the prompt
collapse threshold is primarily determined by the EOS pressure support at
large densities (or the maximum mass). 
For example, for a EOS 
sufficiently soft at the postmerger densities
$\rho\gtrsim3\rho_0$, where $\rho_0$ is the nuclear density, but admitting
small compactness at inspiral densities ($\rho \sim 2\rho_0$), 
Eq.~\eqref{eq:k_thr} might incorrectly predict a NS remnant signal instead of
a prompt collapse.
In practice, we do not have such EOS in our hadronic EOS sample but 
interesting examples are the EOS with hyperons \cite{Banik:2014qja}
or with phase transitions to quark deconfined matter. We will
discuss how to deal with these cases using a specific example below. 
Improvements in the modeling of the prompt collapse threshold and the
waveform amplitudes for the short-lived cases are possible and will be
considered in the near future as more and more accurate simulations will become available. 


\section{Validation of \nrpm{}} 
\label{sec:validation}

\begin{table*}[t]
  \centering    
  \caption{Properties of validation binaries and inference results for
    the subset of postmerger injections. The recovered quantities are
    referred to the minimum SNR required to detect the postmerger signal and
    they correspond to median values and 90\% credible regions.}
  \begin{tabular}{cccccccc|ccccccc}      
    \hline
    \hline
    \multicolumn{8}{c|}{Properties} 
    & \multicolumn{7}{c}{Injections' Recovery} \\
    \hline
    EOS &$M^{\rm TOV}_{\rm max}$&$R^{\rm TOV}_{\rm max}$& $M_A$ & $M_B$ &$\kt2$ & $f_2$& Ref. & SNR$_{\rm MF}$&SNR$_{\rm opt}$& $M$ & $q$ & $\kt2$ & $f_2$ & $R_{\rm max}$  \\
    &  [$\Mo$]&[km] & [$\Mo$]&[$\Mo$]  & & [kHz] &  &(Min.)&(Min.) &[$\Mo$] & &&[kHz] & [km] \\
    \hline
    2B&1.78&8.47&1.35&1.35&23.6&---& \cite{Bernuzzi:2014owa}&---&---&---&---&---&---&---\\
    SLy4& 2.06&9.97&1.364&1.364&75.2&3.65&\cite{Perego:2019adq}&12&22&$1.79^{+0.46}_{-0.17}$&$1.33^{+0.11}_{-0.07}$&$74^{+151}_{-4}$&$5.22^{+0.03}_{-2.30}$&${6.5}^{+3.4}_{-0.3}$\\
    BHB$\Lambda\phi$& 2.10&11.63&1.50&1.50&90.0&3.39&\cite{Radice:2016rys}&10&13&$2.50^{+0.10}_{-0.25}$&$1.03^{+0.07}_{-0.03}$&$79^{+28}_{-8}$&$3.60^{+0.14}_{-0.07}$&${9.3}^{+0.2}_{-0.5}$\\
    DD2 &2.42 &11.93&1.50&1.50&91.1&2.76&\cite{Radice:2016rys}&9&13&$2.39^{+0.35}_{-0.29}$&$1.10^{+0.13}_{-0.09}$& $196^{+79}_{-68}$&$2.74^{+0.02}_{-0.02}$ &${10.6}^{+0.7}_{-0.7}$\\
    SLy4&2.06&9.97&1.30&1.30&93.1&3.13&This work &8 &13&$2.40^{+0.26}_{-0.28}$&$ 1.09^{+0.07}_{-0.08}$&$137^{+54}_{-35}$&$3.11^{+0.02}_{-0.02}$&${9.9}^{+0.5}_{-0.6}$\\
    LS220& 2.04&10.67&1.364&1.364&133.9&2.97&\cite{Perego:2019adq}&8 &13&$2.30^{+2.38}_{-0.44}$&$1.28^{+0.17}_{-0.22}$&$218^{+500}_{-99}$&$2.95^{+0.03}_{-2.07}$&${9.9}^{+7.3}_{-0.7}$\\
    LS220& 2.04&10.67&1.4&1.33&133.9&3.03&This work&9 &14&$2.32^{+0.34}_{-0.25}$&$1.25^{+0.09}_{-0.08}$&$168^{+60}_{-54}$&$3.00^{+0.02}_{-0.02}$&$10.0^{+0.7}_{-0.5}$\\
    DD2&2.42 &11.93&1.364&1.364&157.5&2.39&\cite{Perego:2019adq}&7&12&$1.94^{+2.75}_{-0.43}$&$1.06^{+0.34}_{-0.06}$&$414^{+252}_{-332}$&$2.30^{+0.88}_{-1.42}$&$10.9^{+10.9}_{-5.0}$\\
    H4&2.03&11.66&1.45&1.25&210.7&2.33&\cite{Dietrich:2015iva}&6& 8 & $4.01^{+0.97}_{-2.25}$&$1.27^{+0.19}_{-0.24}$&$183^{+554}_{-107}$&$1.85^{+0.88}_{-0.99}$&$16.8^{+2.3}_{-7.1}$\\
    BHB$\Lambda\phi$& 2.10&11.63&1.25&1.25&256.1&2.36&\cite{Radice:2016rys}&8&9&$2.41^{+0.42}_{-0.26}$&$1.07^{+0.15}_{-0.07}$&$281^{+88}_{95}$&$2.35^{+0.02}_{-0.02}$&${11.5}^{+0.9}_{-0.5}$\\
    \hline
    \hline
  \end{tabular}
  \label{tab:valbns}
\end{table*}

We compare the \nrpm{} model to all non-spinning binaries in {\tt CoRe} database
and to a ``validation set'' of 10 simulations that were not employed for the fits of Sec.~\ref{sbsec:NRinfo}. 
The properties of the validation set are summarized
in Tab.~\ref{tab:valbns}. The simulations span 
the relevant ranges in $\boldsymbol{\theta}$, in particular covering the prompt
collapse and short-/long- lived remnant cases. We compute the mismatch~\cite{Cutler:1994ys}
\be
\MM = 1 -
\max_{\phi_0,t_0}\frac{(h_1(\phi_0,t_0),h_2)}{\sqrt{(h_1,h_1)(h_2,h_2)}} \ ,
\ee
based on the Wigner scalar product between two waveforms 
\be
(h_1,h_2) = 4\Re\int_{f_\text{min}}^{f_\text{max}}
\frac{\tilde{h}^*_1(f)\,\tilde{h}_2(f)}{S_n(f)}\,{\rm d}f\ ,
\ee
and assuming advanced LIGO design sensitivity \cite{TheLIGOScientific:2014jea,Aasi:2013wya,Harry:2010zz}
for the power-spectral-density (PSD) function $S_n(f)$ and $[f_\text{min},f_\text{max}]=[f_\text{mrg},4096\:\text{Hz}]$.
The value of $\MM$ represents the loss in signal-to-noise ratio
(squared) for waveforms that are aligned in time and phase. Additionally, we analyze
time-domain phasing between the model and the NR waveforms.

\begin{figure}[t]
  \centering 
  \includegraphics[width=.49\textwidth]{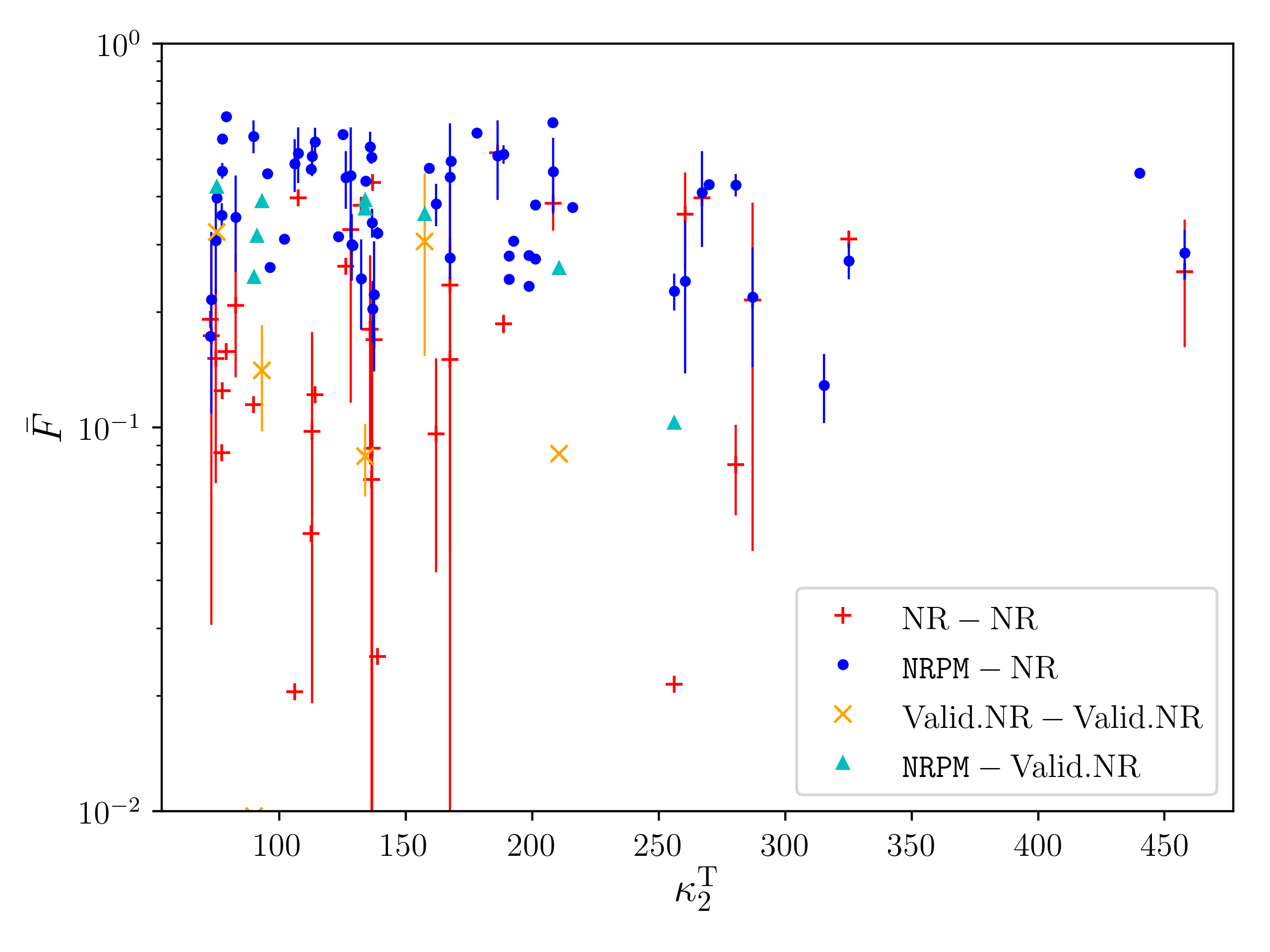}
  \caption{Mismatches between {\tt NRPM} model and the {\tt CoRe} NR
    waveforms. The validation set is indicated with cyan triangle markers.
    Vertical bars indicate the range of mismatches amongst NR
    waveform at different grid resolution (when available); a single
    marker indicates the mismatch between waveforms from two grid
    resolutions or the average from many resolutions.
    LIGO design sensitivity \cite{TheLIGOScientific:2014jea,Aasi:2013wya,Harry:2010zz}
    is used in the calculation of $\MM$ and the 
    frequency ranges start from $f_\text{mrg}$ (computed with relation extracted above) 
    and reach 4kHz.} 
  \label{fig:mm}
\end{figure}

Mismatches against the {\tt CoRe} data used in the fits
are shown in
Fig.~\ref{fig:mm}; the points relative to the validation set waveforms are shown as cyan triangle markers. 
The plot orders the binaries according to $\kt2$.
The largest mismatches are of order ${\sim}0.65$ for
$\kt2\lesssim200$, smallest mismatches are of order ${\sim}{0.1}$, and on average $\MM\sim 0.3$. We recall that a mismatch $\MM$
roughly corresponds to a fractional reduction in detection rate of
${\sim}1-(1-\MM)^3$ for sources that are uniformly distributed in space
\cite{Lindblom:2008cm,Lindblom:2009ux}. Template banks for detection
are usually constructed such that the maximum value of $\MM$ across
the bank is $0.03$, thus allowing for a ${\sim}10$\% loss in the detection rate.
The requirements for parameter estimation are believed to be 
more restrictive than those for detection, but current state-of-the-art
binary-black-hole EOB waveforms have $\MM\sim(0.001-0.01)$, e.g.~\cite{Nagar:2018zoe}.
Mismatches of \nrpm{} with NR waveforms are obviously larger than
those of models that directly use the same NR data
\cite{Clark:2015zxa,Bose:2017jvk,Easter:2018pqy} (Note however less
than 40 simulations were used in those works).  
They are instead comparable to those of \cite{Tsang:2019esi} obtained
with a similar dataset and overall model design. 

The mismatches should also be compared to the NR
uncertainties.
For each binary, we plot an 
estimate of the NR uncertainty obtained by computing the mismatch
between simulations at different resolutions. For most of 
the NR data available it is neither possible to show convergence of the
postmerger waveform phase nor a monotonic
behaviour with grid resolution (but see
\cite{Radice:2016rys,Radice:2016gym} for counter examples and
Appendix~\ref{app:long-conv} for a discussion on error controlled postmerger
waveforms). Hence, we pragmatically compute 
mismatches between waveforms from all the pairs of simulations at the 
different grid resolutions available. 
From Fig.~\ref{fig:mm} it is clear that postmerger NR
data do not satisfy by themselves the $\MM\lesssim0.03$ criterion,
and NR mismatches are in many cases comparable to those due to the modeling.
A necessary condition for the development of faithful postmerger
models is thus the improvement of the NR postmerger waveforms.

\begin{figure*}[t]
  \centering 
  \includegraphics[width=.49\textwidth]{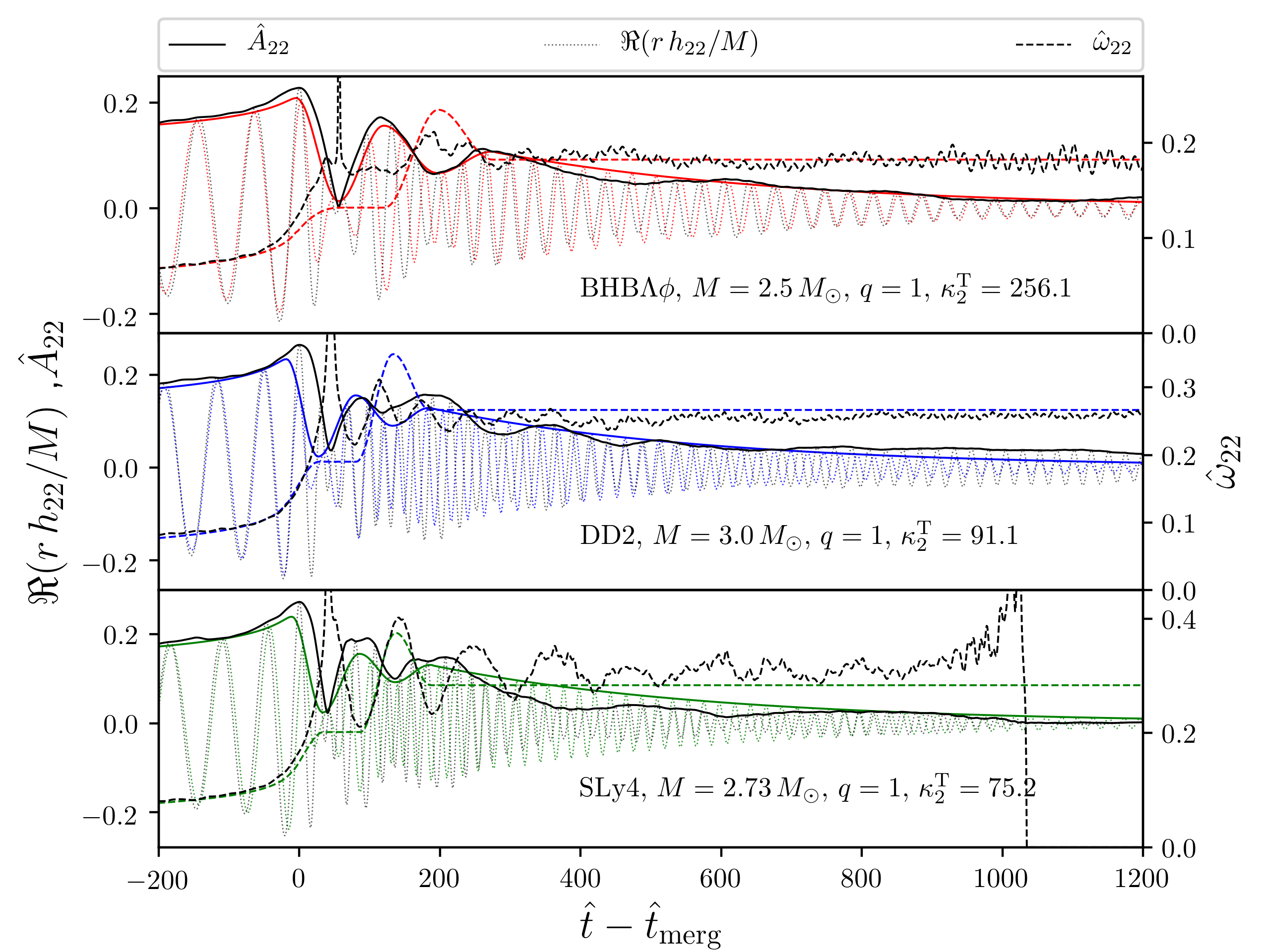}
  \includegraphics[width=.49\textwidth]{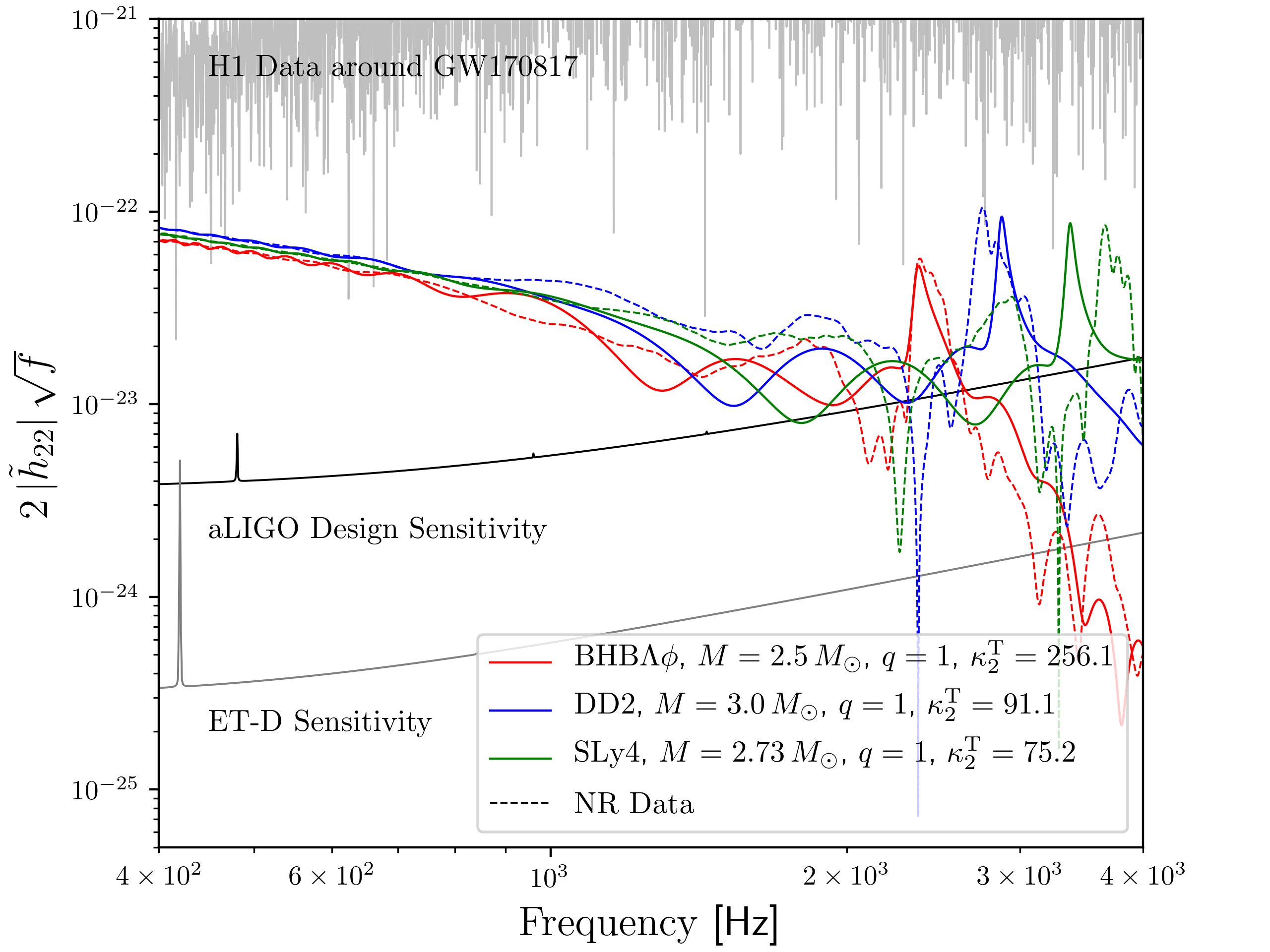}
  \caption{Complete \texttt{TEOBResumS\_NRPM} $(2,2)$ waveforms 
    and corresponding spectra.
    Left panel: Time-domain \texttt{TEOBResumS\_NRPM} $(2,2)$ waveforms 
      compared with selected NR hybrids around merger. From top to bottom, 
      BHB$\Lambda\phi$ $M=(1.25+1.25)\Mo$ is the best mismatch case, 
      DD2 $M=(1.50+1.50)\Mo$ represents an intermediate case and 
      SLy4 $M=(1.364+1.364)\Mo$ is the worst mismatch case. 
      Right panel: Corresponding spectra from 400~Hz to 4~kHz with sources
      located at 40~Mpc and analytical power spectral densities of LIGO design 
      \cite{TheLIGOScientific:2014jea,Aasi:2013wya,Harry:2010zz} 
      and Einstein Telescope \cite{Punturo:2010zz,Hild:2010id}.}
    \label{fig:fullmodel}
\end{figure*}

We further discuss time-domain phasing and spectra for three binaries 
taken from the validation set and shown in  Fig.~\ref{fig:fullmodel}. The best
match case is the BHB$\Lambda\phi$ with $M=(1.25+1.25)\Msun$ ($\MM\sim0.1$) for
which the peak frequency
$f_2=2358$~Hz is well
reproduced by the model (fit value $f_2^\text{fit}=2357$~Hz) and
the waveform remains in phase for 
${\gtrsim}10$~ms after merger. Phase differences at late times influence
less the match since most of the energy is radiated earlier. 
The DD2 with $M=(1.50+1.50)\Mo$ has a moderate match with \nrpm{}. The
model slightly overestimates $\f_2$ predicting
$f_{2}^\text{fit}=2871$~Hz instead of $f_{2}=2761$~Hz. 
Some significant dephasing is observed around $\hat{t}\sim200$ for several cycles, and it is likely the main cause of the mismatch.
The worst mismatch is obtained
with the SLy4 with $M=(1.364+1.364)\Mo$ that produces a short-lived remnant
collapsing in ${\sim}13$~ms. For this BNS the peak frequency is
underestimated by the model ($f_{2}=3654$~Hz vs
$f_{2}^\text{fit}=3367$~Hz). The NR frequency evolution has several oscillations and increases before collapse; these features are not modeled by \nrpm{}. 
Consequently, the model has a poor
match. Note the $\f_{2\pm0}$ are rather well estimated in this case.

Inspection of other waveforms confirms that mantaining the phasing in the early
postmerger signal is a key factor for the overall accuracy of the model.
In addition, since the $\f_2$ fits of Sec.~\ref{sec:NRPM} are
less accurate for small $\kt2$, \nrpm{} better describes the 
waveforms of BNS with larger $\kt2$ corresponding to lower postmerger
frequencies. Note the latter are the most favored in low SNR
detections. In other words, \nrpm{} is more robust (uncertain) for
long-lived (short-lived) remnant, as expected. 
Finally, we test a simpler version of \nrpm{} with the single
frequency $\f_2$ and find that some short-lived data are actually
better described by this simpler model which averages the frequency evolution.


\section{Time-domain inspiral-merger-postmerger model} 
\label{sec:IMPM}

\begin{figure}[t]
  \centering 
  \includegraphics[width=0.49\textwidth]{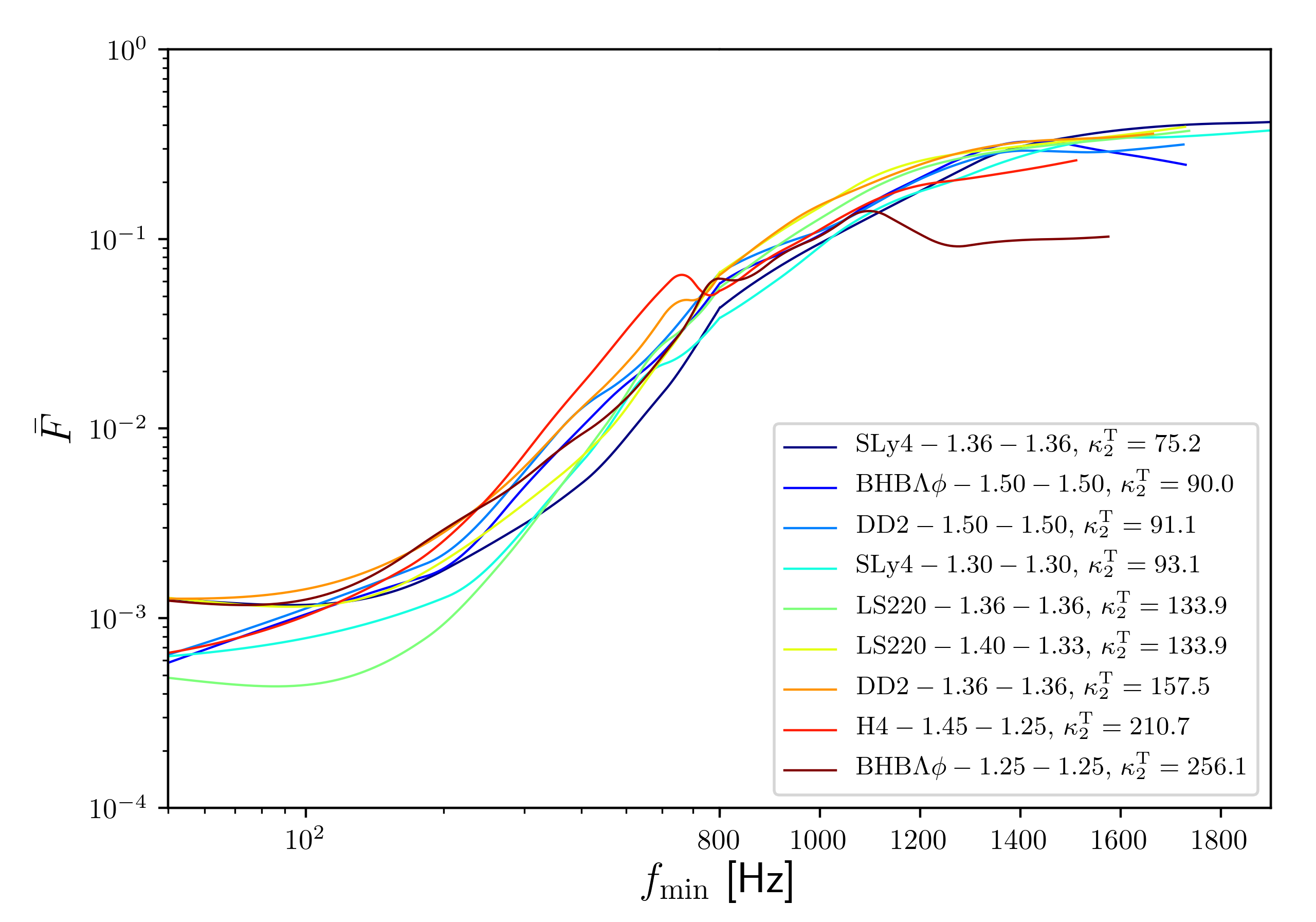}
    \caption{Mismatches between hybrid waveforms (\texttt{TEOBResumS}
      + NR) and the complete model \texttt{TEOBResumS\_NRPM} as function 
      of lower cut-off frequency $f_\text{min}\in [50~{\rm Hz},f_\text{mrg}]$. The latter quantity is taken from the NR fits.}
    \label{fig:mm_fmin}
\end{figure}

A model for the time-domain inspiral-merger-postmerger (IMPM) waveform is 
obtained by smoothly attaching amplitude and 
phase of \nrpm{} at the peak amplitude $\hat{A}_\text{mrg}$ of any time-domain
inspiral-merger model. Currently, the only time-domain waveforms that can
reproduce the merger peak amplitude are the effective-one-body (EOB) ones.
We thus use the tidal EOB model developed in
\cite{Bernuzzi:2014owa,Nagar:2018zoe,Akcay:2018yyh}
and called \texttt{TEOBResumS}. 

The attachment is done at the amplitude peak as described in
Sec.~\ref{sbsec:TDmodel}, but using the amplitude
$\hat{A}_\text{mrg}$, the merger frequency $\w_\text{mrg}$ and its
derivative $\dot{\w}_\text{mrg}$ of the inspiral-merger waveform.
Amplitudes $\hat{A}_i$ are then fixed by computing the ratios $\beta_i$.
Examples of IMPM waveforms are shown in Fig.~\ref{fig:fullmodel} and
compared to NR waveforms. In order to perform a visual comparison, the
NR and \texttt{TEOBResumS\_NRPM} waveforms are aligned in phase and
time at merger. The figure shows the smooth attachment at merger and
the phase coherence of the postmerger completion. 
The figure also highlights that \nrpm{} is more accurate for BNS with
larger $\kt2$, as discussed in Sec.~\ref{sec:validation}. 

A quantitative measurement of the phase coherence is obtained by
computing mismatches between the \texttt{TEOBResumS\_NRPM} model and
hybrid waveforms constructed joining \texttt{TEOBResumS} to NR
data. We built such hybrid waveforms starting from a GW frequency of
50~Hz and for each BNS of the validation set. The mismatches are
computed as functions of the lower cut-off frequency $f_\text{min}$,
which takes values from 50~Hz to $f_\text{mrg}$.
where the latter is obtained by the NR fits. Figure~\ref{fig:mm_fmin}
shows the mismatches as a function of $f_\text{min}$ for the
validation set. Significant phase differences are accumulating between 500~Hz and
800~Hz where the NR merger is attached. The last point of each line
corresponds to the mismatch between \nrpm{} and NR; typical values are
$\MM\lesssim0.3$ with a minimum $\MM\sim0.1$ consistently with what
discussed in Sec.~\ref{sec:validation}.


\section{Injection studies} 
\label{sec:injection}

To demonstrate the applicability of \nrpm{} in the context of Bayesian GW
data analysis we consider a set of experiments in which known signals are 
injected using zero-noise configuration and recovered using standard Bayesian 
inference techniques. The experiments aim at addressing
the following questions:   
\begin{itemize}
\item[(A)] At which SNR can \nrpm{} detect a PM signal? 
\item[(B)] Is it possible to infer whether the merger remnant 
  collapsed to a BH or was a NS using the IMPM model?
\item[(C)] What constraints can be set on the NS 
  minimal radius from the PM analysis solely?  
\item[(D)] Is it possible to infer the EOS stiffness at the 
extreme densities reached in the NS remnant using the IMPM signal?
\end{itemize}

Given data $d$ and hypothesis $H$, the posterior distribution of the
parameters $\boldsymbol{\Theta}$ is defined from Bayes' theorem, 
\be
\label{eq:bayes}
p(\boldsymbol{\Theta}|d, H) =\frac{p(d|\boldsymbol{\Theta},H)\,p(\boldsymbol{\Theta}|H)}{p(d|H)}\ ,
\ee
where $p(\boldsymbol{\Theta}|H)$ is the prior distribution for the parameters
$\boldsymbol{\Theta}$ and 
$p(d|\boldsymbol{\Theta},H)$ is the likelihood function. For a single detector $i$, 
the likelihood is defined as
\be
\log p_i(d|\boldsymbol{\Theta},H) \propto -\frac{1}{2} \left( d-h_{\boldsymbol{\Theta}}  ,
d-h_{\boldsymbol{\Theta}} \right)_i\ ,
\ee
where $h_{\boldsymbol{\Theta}}$ is the GW template, which depends on the parameters $\boldsymbol{\Theta}$. 
For a detector network it is obtained multiplying the likelihood of
the single detectors. 
The term $p(d|H)$ is the evidence and it can be computed as 
the marginalization of the likelihood function over the entire parameters space.

We perform two sets of experiments using the amplitude sensitivity
densities (ASD) of the three Advanced LIGO \cite{TheLIGOScientific:2014jea,Aasi:2013wya,Harry:2010zz} 
and Advanced Virgo detectors \cite{TheVirgo:2014hva}.
In the first set, we inject 9 postmerger signals of the validation set reported in
Tab.~\ref{tab:valbns} placing the source at $2,3,4,5,6,7,8$~Mpc and 
located at right ascension and declination $(\alpha,\delta)=(0,0)$ 
with angle of view $\iota=0$, polarization angle $\psi = 0 $ and sampled at 8192~Hz.
In the injections, we apply a Tukey window at merger in order to isolate the postmerger signal and 
remove the contributions from the inspiral.
The distances approximately correspond to postmerger SNRs from 4 to 16, 
with the exact values depending on the particular BNS.
The injected NR signals are recovered with \nrpm{} by analyzing the
frequencies $[1024,4096]$~Hz and fixing the sky location of the source.
Inference is performed on the extended set of parameters
\be
\boldsymbol{\Theta}=(M_A,M_B,\Lambda_A,\Lambda_B,D_L,\psi,t_0,\phi_0) \ ,
\ee
where $(t_0,\phi_0)$ are the time shift and the merger phase, respectively, and $\psi$ is the polarization angle.
In this paper we prescribe the collapse threshold as
$\kappa^{\rm T}_\text{thr} = 70$; for more general analysis the parameter
can be included into $\boldsymbol{\Theta}$. 
We also use the $\alpha$ parameter in Eq.~\eqref{eq:alpha} as estimated from the
NR fits but, as discussed in Sec.~\ref{sbsec:NRinfo:fat}, uncertainties
on the $\alpha$ fit can lead to incorrect distance estimates.
In future analysis it should be explored the effect of promoting 
$\alpha$ to an inference parameter, effectively allowing for 
a more agnostic analysis.

The posterior distributions of other parameters are recovered using
their definitions or from the fits in case of peak frequencies.
Priors are set on chirp mass, mass ratio and $\Lambda_{A,B}$, that are
bounded to $\M_c\in[0.5,2.2]\Mo$, $q\in[1,1.5]$ and
$\Lambda_{A,B}\in[50,5000]$. The prior distributions are uniform in the
individual components $M_{A,B}$ and $\Lambda_{A,B}$.
Bayesian inference is performed with the nested sampling
algorithm~\cite{Skilling:2006} 
as implemented in the \texttt{LALInference} software
package~\cite{Veitch:2009hd,Veitch:2014wba,lalsuite}. 

In the second set, we inject hybrid waveforms and we recover with
either the IM model or the IMPM model. Specifically, we use the 
nonspinining surrogate of \texttt{TEOBResum} developed in
\cite{Lackey:2016krb} and refer to the IM (IMPM) model as 
\texttt{TEOBResum\_ROM} (\texttt{TEOBResum\_ROM\_NRPM}).
The choice of the priors is identical to the previous cases, except 
for the chirp mass for which we use a smaller range $\M_c\in[1,2.2]\Mo$
and the frequency range analyzed is $[50,4096]$~Hz.
We note that the injection labelled as 2B $M+(1.35+1.35)\Mo$ is a prompt-collapse signal.
\nrpm{} does not include a template for this type of sources and then this waveform is excluded
from the detectability application, but it is included in the second set of injections (Sec.~\ref{sbsec:inferpc}).

Considering a GW170817-like source, an optimal SNR ${\sim}3$ could be achieved by the
Advanced LIGO-Virgo detectors at design sensitivity, while SNR ${\sim}10$ is expected to
be achieved by third generation detectors.
From now on, the SNR value we quote is the maximum value coming from
the matched-filtered SNR computation between \nrpm{} model and the injected signal.

\subsection{Postmerger detectability}
\label{sbsec:pmdetect}

\begin{figure*}[t]
  \centering 
    \includegraphics[width=\textwidth]{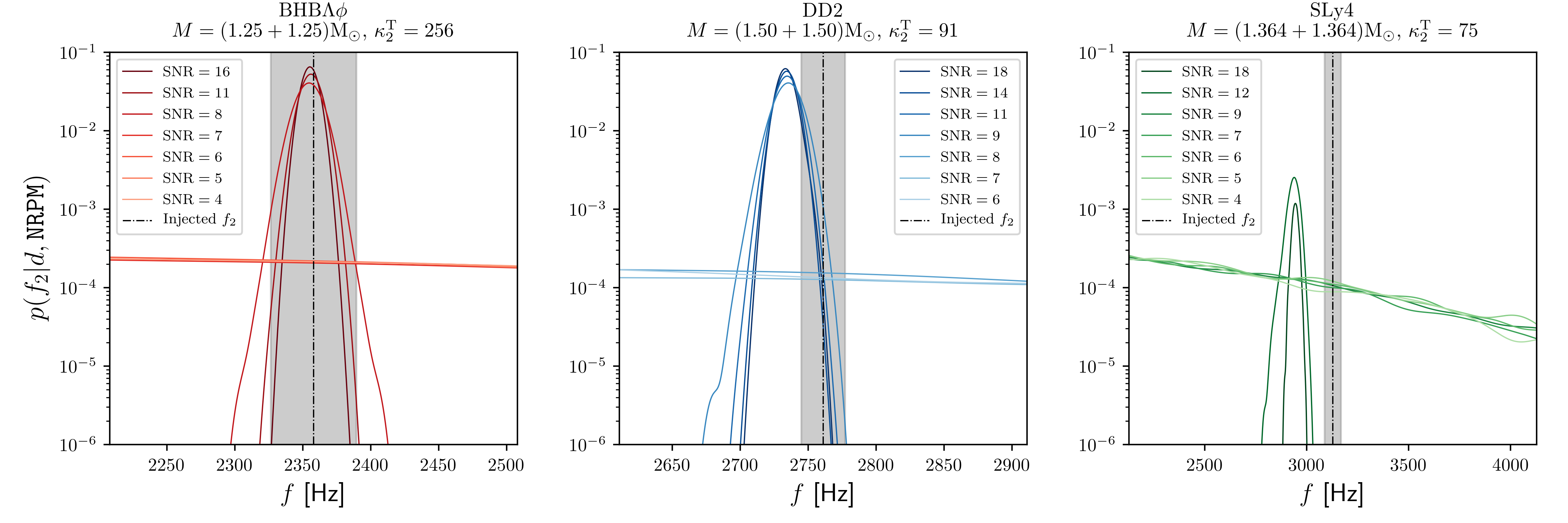}
    \caption{Marginalized posterior distributions of $f_2$ for three injected cases
      at different SNRs: the first case, BHB$\Lambda\phi$ $M=(1.25+1.25)\Mo$, 
    is a case where the peak frequency is well recovered and this is
    also supported by the low mismatch between \nrpm{} model and the
    injected signal.  
    In the second case, DD2 $M=(1.50+1.50)\Mo$, we can see that for
    high SNRs biases appear systematically and the recovered peak is
    below the injected one. 
    The third case, SLy4 $M=(1.364+1.364)\Mo$, shows a bimodal
    distribution: a dominant peak appears at frequency ${\sim}5.2$~kHz
    (beyond the Nyquist limit, not in the plot) while the secondary peak is close to the
    injected value. The primary peak is compatible with the
    frequency $f_{2-0}$ aliased at high frequencies.}      
    \label{fig:posterior}
 \end{figure*}

We discuss the results of the first set of injections employing only
PM signals and \nrpm{}.
The matched filtering analysis of the validation set gives evidence of postmerger
signals starting from network SNR $\sim8-9$.
The latter correspond to source distances of ${4-6}$~Mpc.
We find that statistical errors are larger than systematic
uncertainties at SNR ${\lesssim}12$ but the two become comparable for higher SNRs.

The parameters recovered by the analysis at the minimal SNR are
reported in Tab.~\ref{tab:valbns}.  
For most of the cases, the posterior distributions of the physical
parameters include the injected values within the ${95}$~\%
confidence regions.However, some cases show degeneracies among the
model's parameters.
In general, the largest discrepancies in the recovered parameters are
induced by the inaccuracy of the NR frequency fit for the particular BNS. 
The posterior distributions for $f_2$ for three exemplary cases at
different SNRs are shown in Fig.~\ref{fig:posterior}.
\nrpm{} recovers the correct peak frequency within the uncertainties 
for all the injected binaries except for the DD2 $M=(1.50+1.50)\Mo$ which
will be discussed in the next Sec.~\ref{sbsec:infereos}.

For the injection BHB$\Lambda\phi$ $M=(1.25+1.25)\Mo$, the estimation of the parameters 
with \nrpm{} is in agreement with the injected properties. The posterior
distributions are unimodal and centered around the injected value. In this case, the model is able to 
reconstruct the spectrum of the signal and this fact is also motivated by the low mismatch
between this waveform and the model.

A difficult case is SLy4 $M=(1.364+1.364)\Mo$ for which the value of the masses
and $\kt2$ are underestimated to compensate the smaller
values of $\f_2$ estimated from the NR fits, and to obtain a signal matching the injection ($f_2 \propto M^{-1}$).
Moreover, the marginalized posterior distribution of $f_2$ has a bimodality.
For this signal, $f_2$ is at the edge of the frequency range where 
the sensitivity is smaller and the recovery with \nrpm{} promotes the subdominant peak
$f_{2-0}$ as main frequency, especially for high SNR. However, the $f_{2-0}$ 
is aliased to high frequencies and the maximum of the marginalized posterior distribution 
of $f_2$ is well above the Nyquist frequency of ${\sim}4$~kHz (not shown in the plot). The secondary maximum of the distribution is compatible with the injected value within the uncertainties.  

Another interesting case is BHB$\Lambda\phi$ $M=(1.50+1.50)\Mo$: this postmerger signal
is very short and the remnant collapse after ${\sim}3$~ms. As consequence, the frequency evolution is not trivial
and none of the spectrum peaks is relevantly dominant, since the remnant evolves towards collapse.
Then, the recovered $f_2$ peak is overestimated while the $f_{2-0}$ peak 
is correctly captures ($f_{2-0}^{\rm inj}=2.48$~kHz vs $f_{2-0}^{\rm rec}=2535^{+40}_{-48}$~Hz at SNR 11).


In general, we observe for some cases a shift in the recovered value of the total mass $M$:
this parameter strongly correlates with the position of the frequency peak and with its amplitude in the 
frequency domain. The latter quantities are also determined by the damping time in Eq.~\eqref{eq:alpha},
whose behavior is not well capture by NR fits (Tab.~\ref{tab:fitcoefs}). These uncertainties propagate
during the parameter estimation routine and the results are biased. However, these effects could be 
avoided including $\alpha$ into $\boldsymbol{\Theta}$. Moreover, this estimation can be 
inferred with high accuracy from the inspiral measurement at these SNRs.

\subsection{Inferring prompt-collapse}
\label{sbsec:inferpc}

We discuss the results of the second injection set focusing on two different BNS:
2B $M=(1.35+1.35)\Mo$ which end in a prompt collapse, and BHB$\Lambda\phi$
$M=(1.25+1.25)\Mo$ for which the outcome is a long-lived remnant (see Fig.~\ref{fig:fullmodel}).
In the context of Bayesian analysis, a natural approach for
prompt-collapse inference is to perform model selection between
inspiral-merger and inspiral-merger-postmerger models for given data.
In case of prompt collapse, the IM model should be favored with respect
to the IMPM one, while in case of a long-lived NS remnant it should be the opposite.
Note this analysis relies on the existence of a coherent model for
the full spectrum (modeling the IMPM phases), as the one proposed here.

Specifically, we perform model selection using the Bayes' factor $\B$, 
which quantifies the agreement of two different competitive
hypotheses, $H_A$ and $H_B$, with the data. The Bayes' factor is defined as the ratio of the two posterior 
probabilities, however it is possible to prove that it can be computed as the ratio
of the evidences,
\be
\B^{A}_{B} = \frac{p(d|H_A)}{p(d|H_B)}\,.
\ee
If $\B^{A}_{B} > 1 \,(<1)$, the hypothesis $A\,(B)$ is favored.
In our case, the competitive models are {\tt TEOBResum\_ROM} for the
IM, and {\tt TEOBResum\_ROM\_NRPM} for the IMPM. 
For this test we remove the constraint given by $\kappa^{\rm T}_\text{thr}$ on {\tt NRPM}. 

\begin{table}[t]
  \centering    
  \caption{Evidences computed 
    for the prompt-collapse inference. The uncertainties are
    estimated with the criterion introduced in Ref.~\cite{Skilling:2006}.
    The label `noise' is referred to the template identically equal to zero.}
    \resizebox{.48\textwidth}{!}{
  \begin{tabular}{cccc}      
    \hline
    \hline
Injection & $\log\B^{\rm IM}_{\rm noise}$ & $\log\B^{\rm IMPM}_{\rm noise}$ & $\log\B^{\rm IMPM}_{\rm IM}$\\
    \hline
2B $M=(1.35+1.35)\Mo$ & $124845^{+1}_{-1}$& $124775^{+1}_{-1}$ & $-70^{+2}_{-2}$\\
BHB$\Lambda\phi$ $M=(1.25+1.25)\Mo$ & $107116^{+1}_{-1}$& $107306^{+1}_{-1}$ & $190^{+2}_{-2}$\\
    \hline
    \hline
  \end{tabular}}
  \label{tab:bayespc}
\end{table}

We inject the 2B and BHB$\Lambda\phi$ signals using an SNR${\sim}$12, sufficient to
detect the postmerger signal with \nrpm{}. We
recover with and without attaching \nrpm{} model at merger.
The values of the Bayes' factors obtained are 
reported in Tab.~\ref{tab:bayespc}. The algorithm is able to distinguish
whether the remnant has  undergone prompt collapse or not: the Bayes' factor
for 2B $M=(1.35+1.35)\Mo$ correctly favors the model 
without postmerger ($\log\B^{\rm IMPM}_{\rm IM} = -70^{+2}_{-2}$). 
Similarly, for BHB$\Lambda\phi$ $M=(1.25+1.25)\Mo$ the presence 
of postmerger signal is favored with respect to the prompt collapse case 
($\log\B^{\rm IMPM}_{\rm IM}=190^{+2}_{-2}$).

We point out that numerical relativity simulations indicate 
that in prompt collapse waveforms a
signal, not described by EOB waveforms, is present after the
amplitude peak. We find that the SNR contribution
of this short, ${\lesssim}2$~ms, postmerger signal in the full
spectrum of 2B $M=(1.35+1.35)\Mo$ is below 4\%. 

\subsection{Constraints on NS minimal radius}

As shown in Tab.~\ref{tab:valbns}, at the minimal SNR the inference on $f_2$ delivers a
result accurate at $2-16$\% (two-sigma).  
Using the EOS-independent relation of  
$f_2(R_{1.6})$ from \cite{Bauswein:2015vxa}, this measurement could be
translated into an estimate of the radius of a nonrotating equilibrium
star of mass $1.6\Mo$ ($R_{1.6}$) with an uncertainty of
${\sim}$1.5~km. In a real scenario this is not particularly
interesting since the radius (or equivalently the tidal parameters,
$R\sim\tilde\Lambda^{1/5}$ \cite{Lackey:2011vz,Read:2013zra}) will be
known with an accuracy at least 
100 times better from the inspiral-merger analysis. We find from our
runs that inspiral-merger inference at the minimal postmerger SNR
delivers $\delta\tilde\Lambda/\Lambda\sim0.04$ and $\delta R/R\sim0.008$.

\begin{figure}[t]
  \centering 
    \includegraphics[width=.49\textwidth]{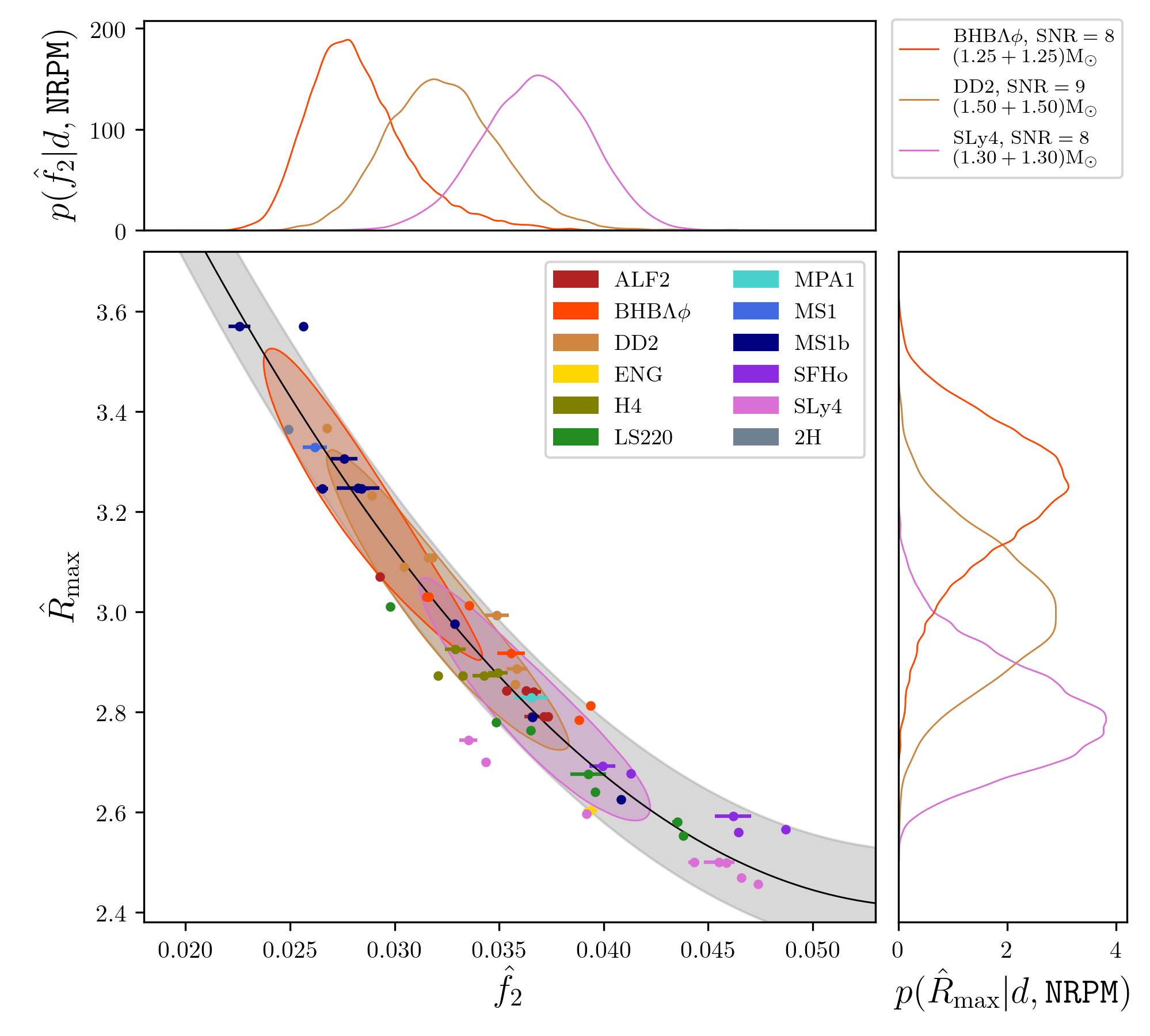}
    \caption{Characteristic postmerger frequency $\hat f_2$ against $\hat R_{\rm max}$ 
    extracted from NR data for different EOS. The black solid line represents the fit with its
    90\% credible region. Right panel shows the marginal posterior distributions of 
    $\hat f_2$ for three selected injections while the top panel shows the 
    respective $\hat R_{\rm max}$ marginal distributions.}
    \label{fig:f2rmax}
 \end{figure}

More interesting is to explore constraints on the radius of the
maximum mass (most compact) nonrotating equilibrium NS $\Rmax$ 
\cite{Bauswein:2014qla}, since the latter corresponds to the largest
matter densities that can be reached for a given EOS.
Using the \texttt{CoRe} NR data, we find an approximate relation in
the form 
\be\label{eq:f2Rmax}
\hat{R}_{\rm max}(\hat{f}_{2}) = (5.81\pm0.13)
-(123.4\pm7.2)\hat{f}_{2}
+(1121\pm99)\hat{f}_{2}^2 \ ,
\ee
where $\hat{R}_\text{max} = \Rmax/M$ and fitting $\chi^2=7.4\times10^{-5}$.
Measurements of PM signals at the minimum SNR deliver an estimation of
$R_\text{max}$ accurate at the ${\sim}$8\% level. The fit
uncertainty is smaller than statistical error at SNR 8, and they become comparable 
for SNR 11.
Figure~\ref{fig:f2rmax} show the data and fit for
Eq.~\eqref{eq:f2Rmax} together with examples of the the posteriors for
$\Rmax$. The latter can be inferred with an uncertainty of
${\sim}1$km. 

Some cases show biased results: for DD2 $M=(1.50+1.50)\Mo$ the expected maximum radius
underestimates the $R_{\rm max}^{\rm TOV}$ predicted by the related EOS,
while for H4 $M=(1.45+1.25)\Mo$ the recovery overestimates the relative value. 
This shifts are coherent with the erroneous estimation
of the total mass $M$, previously discussed in Sec.~\ref{sbsec:pmdetect}.

\subsection{Inferring EOS stiffness at extreme densities}
\label{sbsec:infereos}

\begin{figure}[t]
  \centering 
    \includegraphics[width=.49\textwidth]{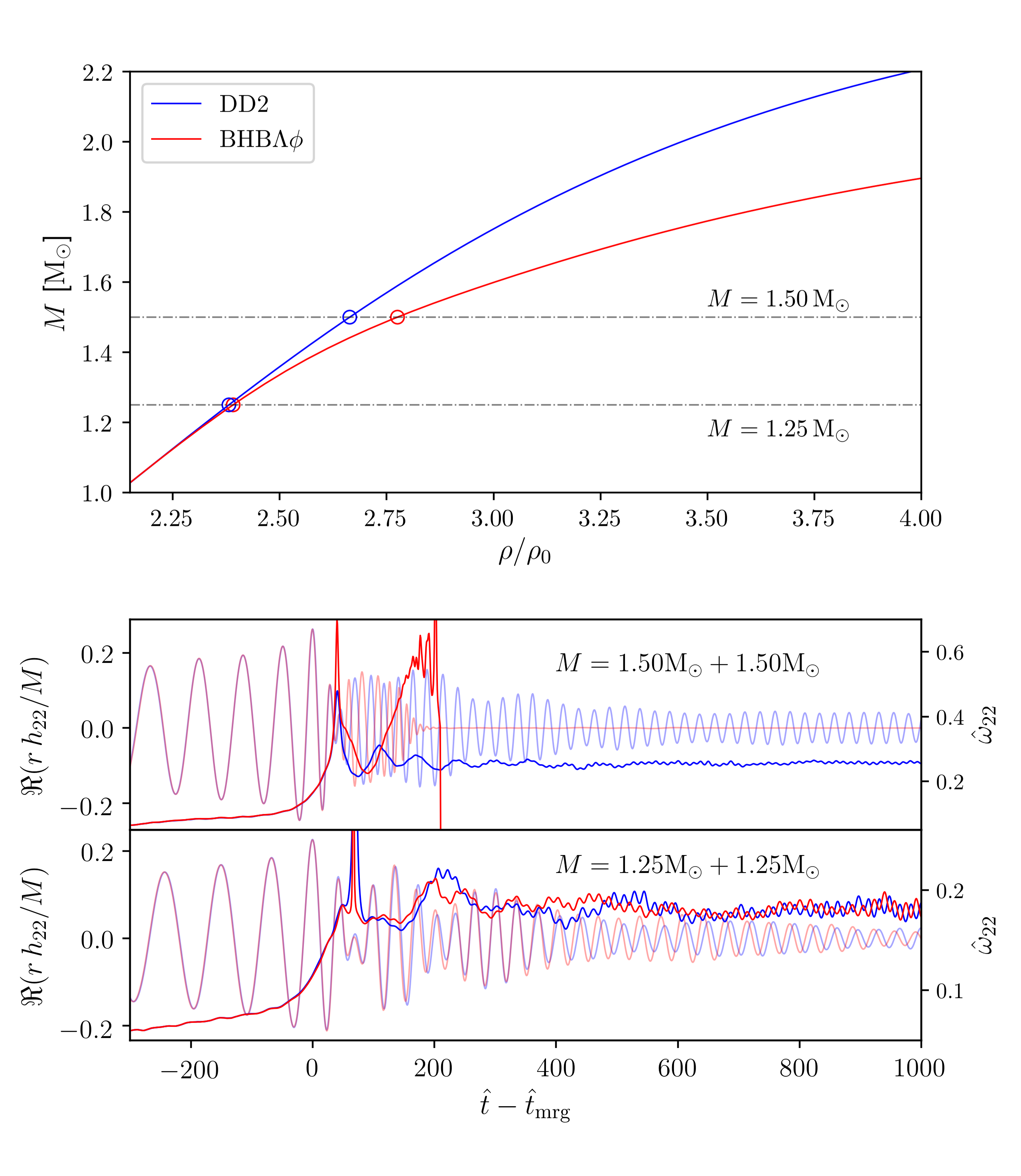}
    \caption{Binary neutron stars described by the BHB$\Lambda\phi$
      and the DD2 EOS and simulated signals~\cite{Radice:2016rys}.
      Top: Mass of individual spherical equilibrium NS as a function
      of the central density. Markers refer to simulated BNS.
      Bottom: Real part of the $(2,2)$ waveforms for 
      BNSs with mass $M=(1.50+1.50)\Mo$ and $M=(1.25+1.25)\Mo$.}
    \label{fig:stiffness}
\end{figure}

We demonstrate the possibility of investigating the EOS 
stiffness at extreme densities using the postmerger GW observations
and \nrpm{}.
We discuss the specific case of EOS
BHB$\Lambda\phi$ and DD2, previously simulated by some of the authors
\cite{Radice:2016rys}.
The BHB$\Lambda\phi$ EOS is identical to DD2 except that at densities
$\rho\gtrsim2.5\rho_0$ (where $\rho_0$ is the nuclear
density) it softens due to the formation of $\Lambda$-hyperons.
Inspiral-merger GW signals from binaries described by the two EOS and
$M\lesssim3\Msun$ are indistinguishable since the
individual NSs have maximal densities $\rho\lesssim2.5\rho_0$, 
similar compactnesses and tidal parameters (same $\kt2$,
Fig.~\ref{fig:stiffness}).

We consider two pairs of binaries: a ``low mass'' with $M=2.5\Msun$
pair and ``high mass'' with $M=3\Msun$ pair.
The individual NS of the low mass BNS have central density
$\rho\approx2.35\rho_0$ and there are essentially no $\Lambda$-hyperons at
these densities in the BHB$\Lambda\phi$ EOS.
The BNS remnants relative to the latter EOS reach approximately
$\rho\approx 2.80\rho_0$ at which BHB$\Lambda\phi$ differs from the DD2 EOS.
The GW postmerger signals have very similar $f_2$ frequencies, but  
they are in principle distinguishable at sufficiently high
SNR \cite{Radice:2016rys}.
The individual NS of the high mass BNS have $\rho\approx2.75\rho_0$;
the presence of $\Lambda$-hyperons significantly affect the postmerger
dynamics. 
The DD2 binary produces
a remnant surviving for ${\gtrsim}20$~ms while the BHB$\Lambda\phi$
binary collapse within ${\sim}2$~ms as a result of the EOS
softening. The postmerger signals are consequently very 
different, as illustrated in Fig.~\ref{fig:stiffness} (bottom panel).


\begin{figure*}[t]
 \centering 
 \includegraphics[width=.49\textwidth]{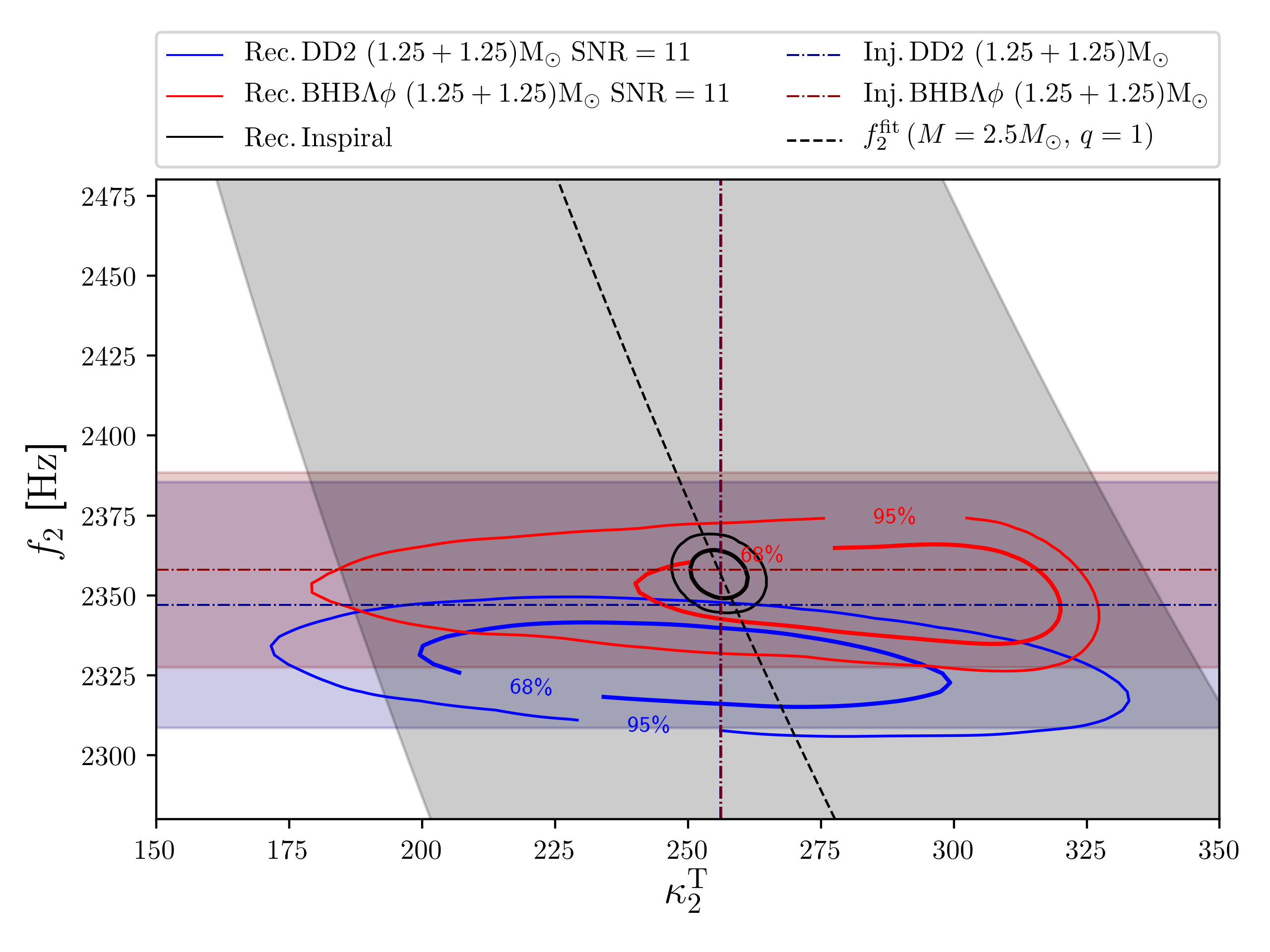}
 \includegraphics[width=.49\textwidth]{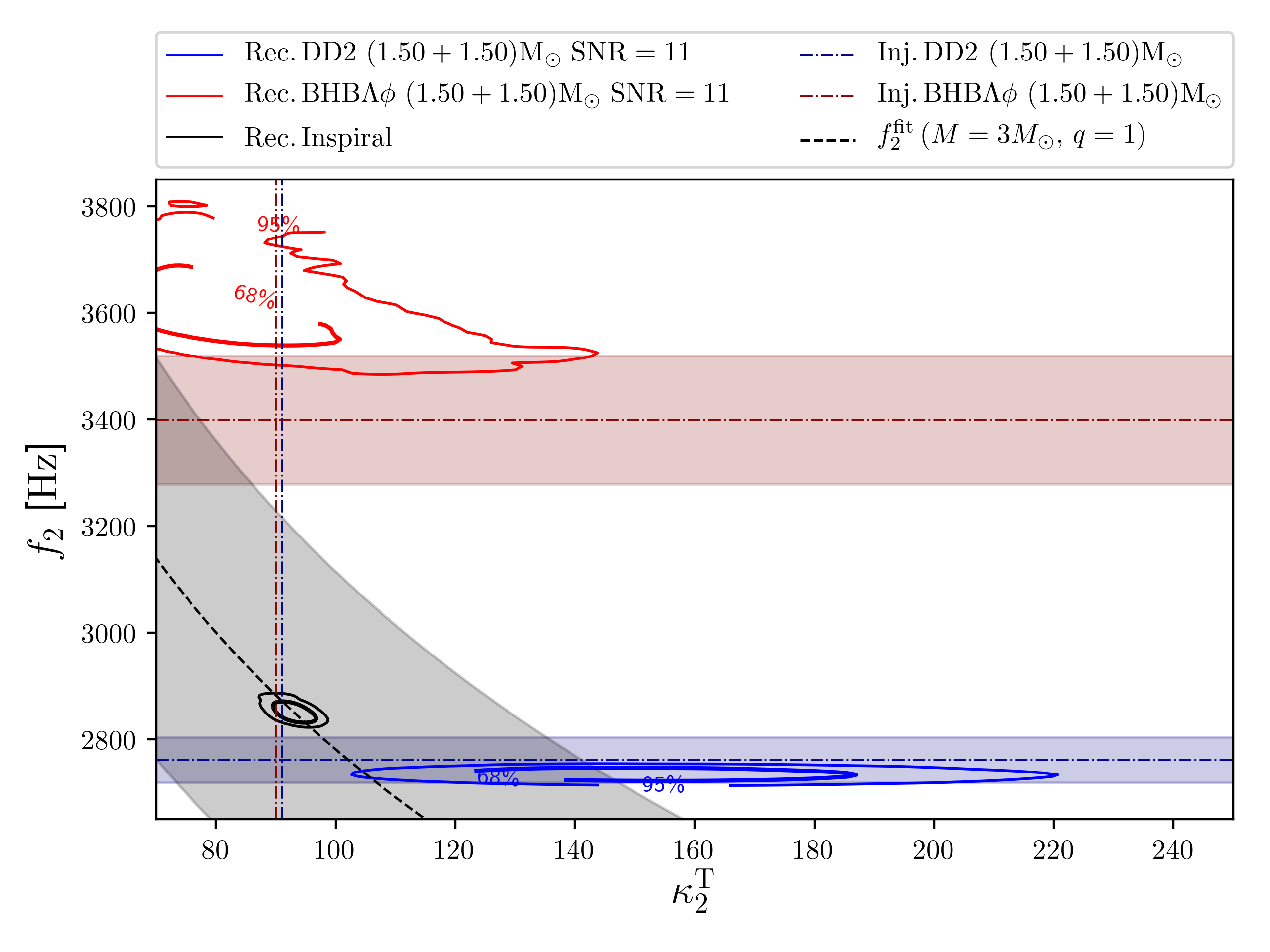}
  \includegraphics[width=.49\textwidth]{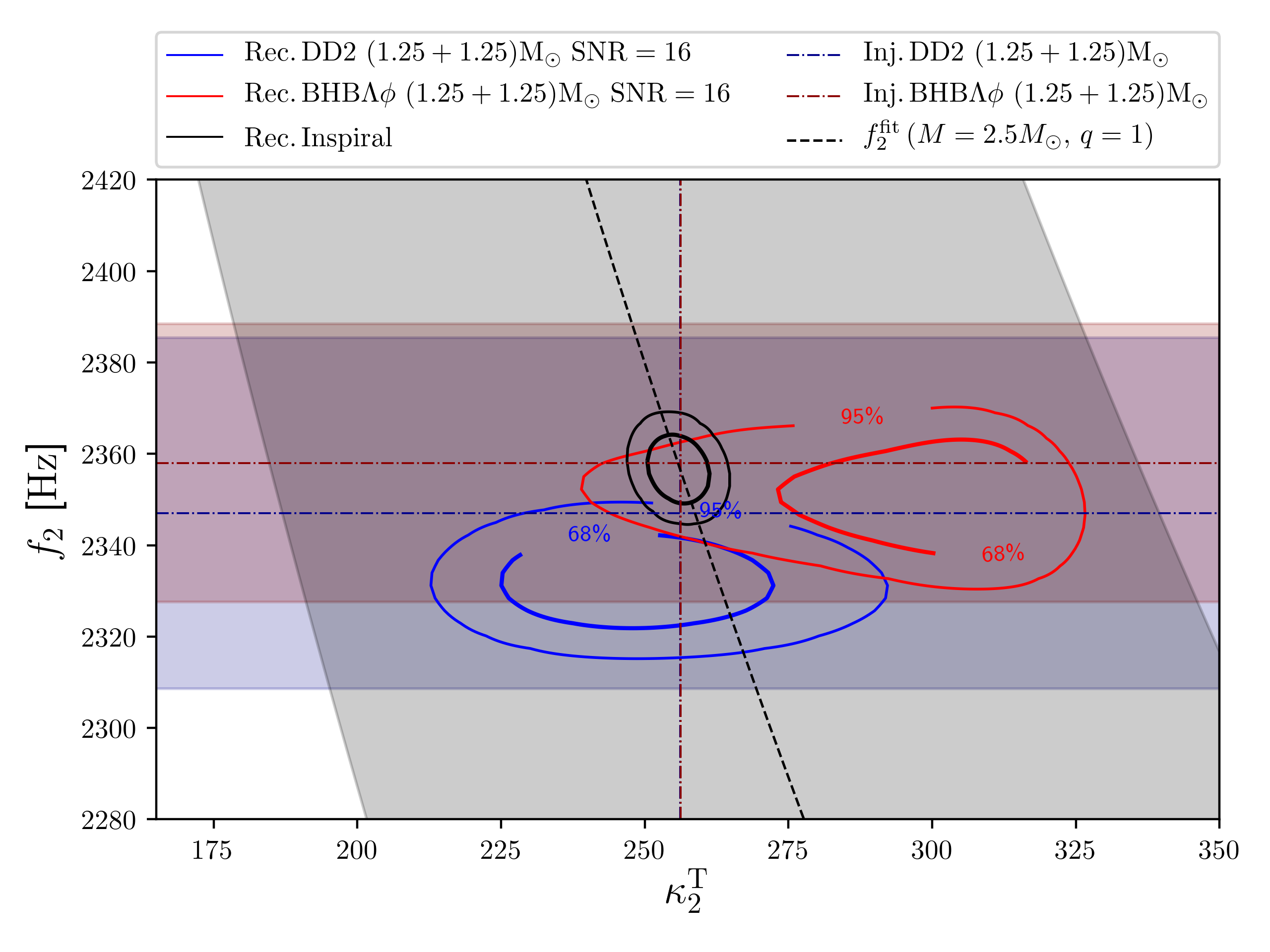}
   \includegraphics[width=.49\textwidth]{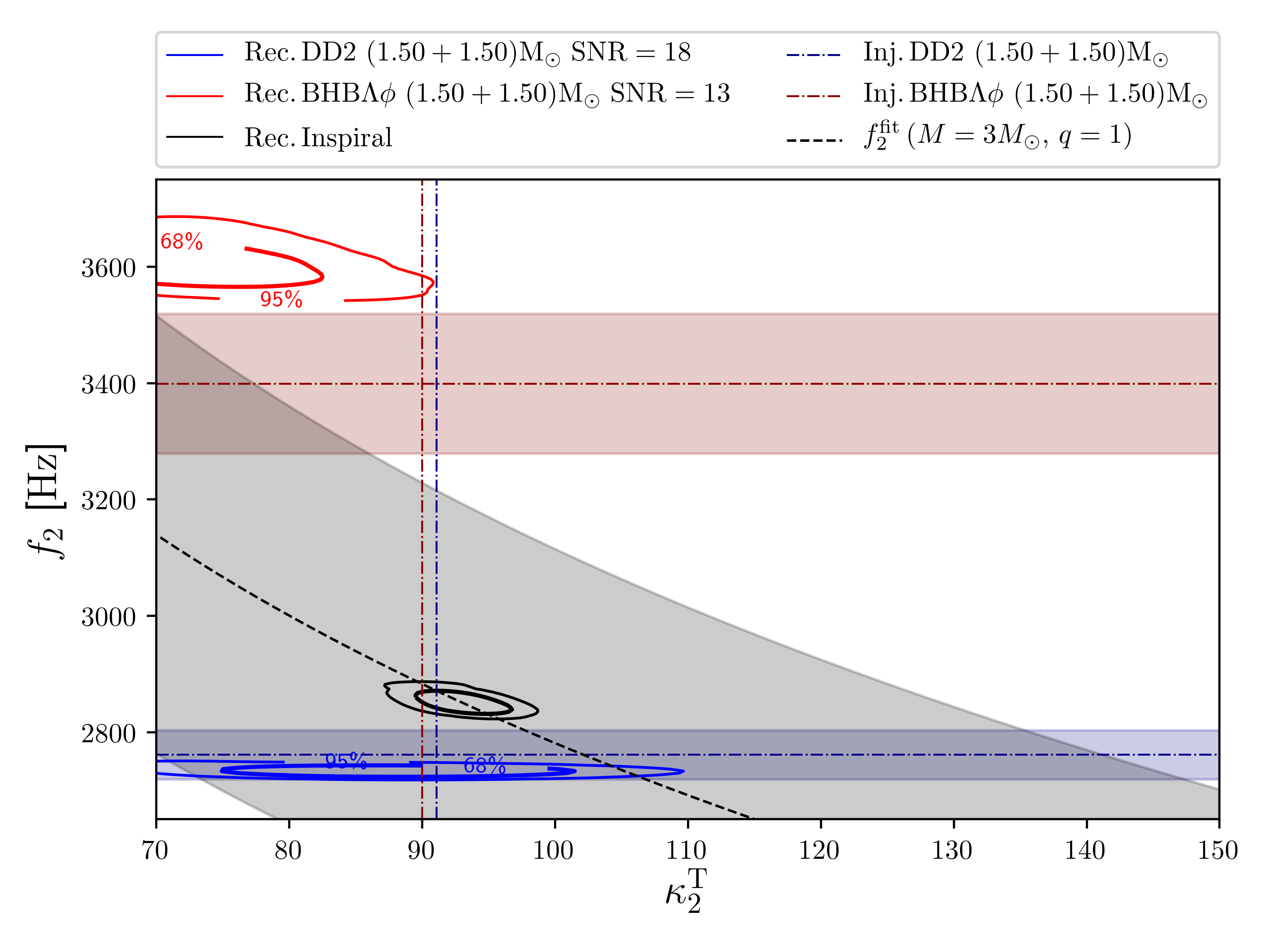}
   \caption{Inference of EOS properties at extreme densities.
    Left panel: marginalized posterior distributions of $f_2$ and $\kt2$ for the 
    ``low mass'' cases (SNR 11 and 16). The postmerger posteriors agree with the value
    predicted by the fit and with the measurement from the inspiral.
    Right panel: marginalized posterior distributions of $f_2$ and $\kt2$ for the 
    ``high mass'' cases (SNR 11 and higher).
    The panels also shows $f_2(\kt2)$ fits related 
    to the injected values with the associated 90\% credible regions.  
    The uncertainties associated to the injected $f_2$ are the widths of the relative peaks
    in the frequency domain.
    }  
   \label{fig:consistency}
\end{figure*}

Figure~\ref{fig:consistency} shows 68\% and 95\% confidence regions of the 
marginal posterior distributions in the $(f_2,\kt2)$ plane as summary
plot of the inference results at two different SNR; the left panels
refers to the low mass BNSs, right panels to high masses. 
The postmerger analysis of the low mass BNSs
returns the injected values and it agrees with the inference from the
inspiral analysis. At SNR 16 some deviations are visible in the
posteriors distribution indicating that such small differences might
be detectable with more accurate models and measurements.

The postmerger analysis of the high mass DD2 $M=(1.50+1.50)\Mo$ shows
that the injected frequency is correctly captured by the recovery,
while the frequency estimated from the inspiral-merger analysis and
the fit is slightly overestimated (as expected, Cf. Fig.~\ref{fig:fullmodel}).
As a consequence of this, the $\kt2$ posterior from the postmerger
analysis is not compatible with the inspiral measurement at the
minimal SNR (upper right panel). However, at higher SNR the correct
$\kt2$ is consistently recovered within the 68\% confidence region
(lower right panel). 

For the BHB$\Lambda\phi$ high mass $M=(1.50+1.50)\Mo$ case, we find
instead inconsistencies between $\kt2$ and $f_2$ posteriors computed
from the IM and PM analysis respectively. 
The postmerger analysis return a $f_2$ higher than the injected
signal, especially at high SNR. At the same time, the $\kt2$
distribution from the postmerger analysis if shifted towards lower values
at larger SNR and rails against the prompt-collapse value
$\kt2\sim70$, significantly departing from the inspiral measurement
$\kappa^\text{T}_\text{2\ IM}={93^{+2}_{-3}}$. The templated-analysis
of the postmerger clearly tries to fit the higher frequencies of the
signal ($f_2=3.39~{\rm kHz}$) and the short postmerger signal
collapsing to BH. The high frequencies of the BHB$\Lambda\phi$ binary 
are incompatible with the quasiuniversal of the \nrpm{} model, due the
physical softening of the EOS. 
Thus, the analysis the postmerger signal effectively implies a softer 
EOS then the analysis of the inspiral implies.

In a real GW measurement the difference in the inferences of $\kt2$ 
(PM vs IMPM results in the high-mass BHB$\Lambda\phi$ case) will give an indication of
the EOS softening at densities larger than those of the individual NS. The
constraint follows from the breaking of the quasiuniversal relation
$f_2(\kt2)$, but the latter does not necessarily imply the presence of new degrees of
freedom or phase transitions (Cf. \cite{Bauswein:2018bma}).
The case studies suggest that a measurement at SNR ${\gtrsim}11$ leads to deviations
from the expected values larger than the 90\% credible regions, 
which is sufficient to make a prediction with significance greater
than one-sigma level.


\section{Conclusion} 

\nrpm{} is a time-domain analytical model for postmerger
waveforms with minimal, but  physically motivated, parameters
describing the morphology of the postmerger waveforms in the binary 
(intrinsic) parameter space defined by Eq.~\eqref{eq:theta}. Combined
with inspiral-merger effective-one-body waveforms, it forms an
approximant coherent in phase on the full frequency range observed by
ground-based interferometers.
Future directions in the modeling of postmerger waveform will include
the extension of the {\tt CoRe} database and the
application of statistical/data reduction methods for the
construction of more accurate and reliable templates\cite{Clark:2015zxa,Easter:2018pqy}.
Central goals for numerical simulations are a better characterization
of the prompt collapse threshold and error-controlled postmerger
waveforms with microphysical EOS and unequal masses.

The current accuracy of the model seems sufficient for the recovery of
signals with postmerger SNR ${\sim}8.5$.
These results, although for a limited set of injections, suggest that 
Bayesian template-based analyses of the postmerger require
higher SNRs than morphology independent analysis
\cite{Chatziioannou:2017ixj,Torres-Rivas:2018svp}.
The latter references claim that about 90\% of the signal can be
reconstructed at SNR ${\sim}5$.
Although a direct comparison of a detectability threshold in the two
types of methods is difficult, the apparent higher requirement in SNR
of the template-based methods is unsurprising, since the latter attempt to
model and recover the entire postmerger signal, as opposed to only
capturing its dominant feature. 
Additionally, the uncertainties associated to numerical relativity simulations 
and to the related fits significantly contribute in the mismatch
(averaging to $\MM\sim0.3$, Fig.~\ref{fig:mm}) and therefore affect
the detectability in the template-based method. 
An advantage of our method is the possibility of performing coherent
analysis of the inspiral-merger-postmerger spectrum. 
We showed that a straightforward application of our models in the
context of Bayesian model selection is the inference of prompt
collapse/remnant star scenarios. 

The quasiuniversal (approximately EOS independent) relations established in this paper
extend previous results and can be employed also with other modeling techniques.
On the one hand, they are key to build waveform models because they
connect the main signal's features with the binary (progenitors NS)
properties.
On the other hand, their direct use to constraining the EOS is not
always relevant.
GW measurements of $R_{1.6}$ or $\kappa_2^\text{T}$ from $f_2$ will not add 
significantly new information on the EOS at extreme densities because 
the inspiral signals of the same sources will deliver more accurate
measurements (stronger EOS constraints) of the same quantities.
For example, the NS radius at fiducial masses would be known at
${\lesssim}10$ meters precision from inspiral measurements against the kilometer
precision of postmerger measurement, with the meter precision being more
accurate than any quasiuniversal relation known to date. 

With this in mind, we have explored a recalibration
[Eq.~\eqref{eq:f2Rmax}] of the relation
$\Rmax(f_2)$ connecting the peak frequency to the radius
of the most compact NS \cite{Bauswein:2014qla}. The latter effectively 
corresponds to the maximal NS central densities, and it is unlikely that
such NS will be components of a binary system.
A single postmerger signal at minimal SNR would deliver $\Rmax$ within
error of ${\sim}8\%$ (few kilometers). Assuming no systematic
effect from the template-based inference, the uncertainty on
$\Rmax$ at minimal SNRs are comparable.


A second constraint of the EOS at extreme densities could come from
the identification of softness effects. We demonstrated that
inconsistencies in the tidal polarizability and in the characteritsic
frequency peak inferred independently from the inspiral-merger and
postmerger analysis can indicate EOS stiffening/softening at densities
${\sim}3-5\rho_0$ already at minimal SNR for detection. 
Note this approach has similarities to the inspiral-merger-ringdown
consistency tests performed on BHs signals
\cite{Ghosh:2016qgn,TheLIGOScientific:2016src,LIGOScientific:2019fpa,Breschi:2019wki}.
It is important to stress that no specific physical
mechanism determining the softening/stiffening is modeled in \nrpm{}
(nor in the NR relations), but the information follows from the breaking
of the specific quasiuniversal relation. 
An interesting development would be to perform model selection on
different postmerger models, should NR quasiuniversal models based on
specific EOS parameterization/families become available.

\begin{acknowledgments}
  The authors thank the LIGO-Virgo matter and postmerger group for
  discussions.
  MB, SB, FZ acknowledge support by the EU H2020 under ERC Starting
  Grant, no.~BinGraSp-714626.  
  DR acknowledges support from a Frank and Peggy Taplin Membership at the
  Institute for Advanced Study and the
  Max-Planck/Princeton Center (MPPC) for Plasma Physics (NSF PHY-1804048).
  Parameter estimation was performed on Virgo ``Tullio'' server 
  at Torino supported by INFN and on LIGO Laboratory supercomputers,
  supported by NSF PHY-0757058 and PHY-0823459.
  Numerical relativity simulations were performed 
  on the supercomputer SuperMUC at the LRZ Munich (Gauss project
  pn56zo), 
  on supercomputer Marconi at CINECA (ISCRA-B project number HP10B2PL6K
  and HP10BMHFQQ), on the supercomputers Bridges, Comet, and Stampede
  (NSF XSEDE allocation TG-PHY160025), on NSF/NCSA Blue Waters (NSF
  AWD-1811236), on ARA cluster at Jena FSU.
\end{acknowledgments}

\appendix


\section{Quasiuniversal relations} 
\label{app:quni-rel}

\begin{figure*}[t]
  \centering 
  \includegraphics[width=.49\textwidth]{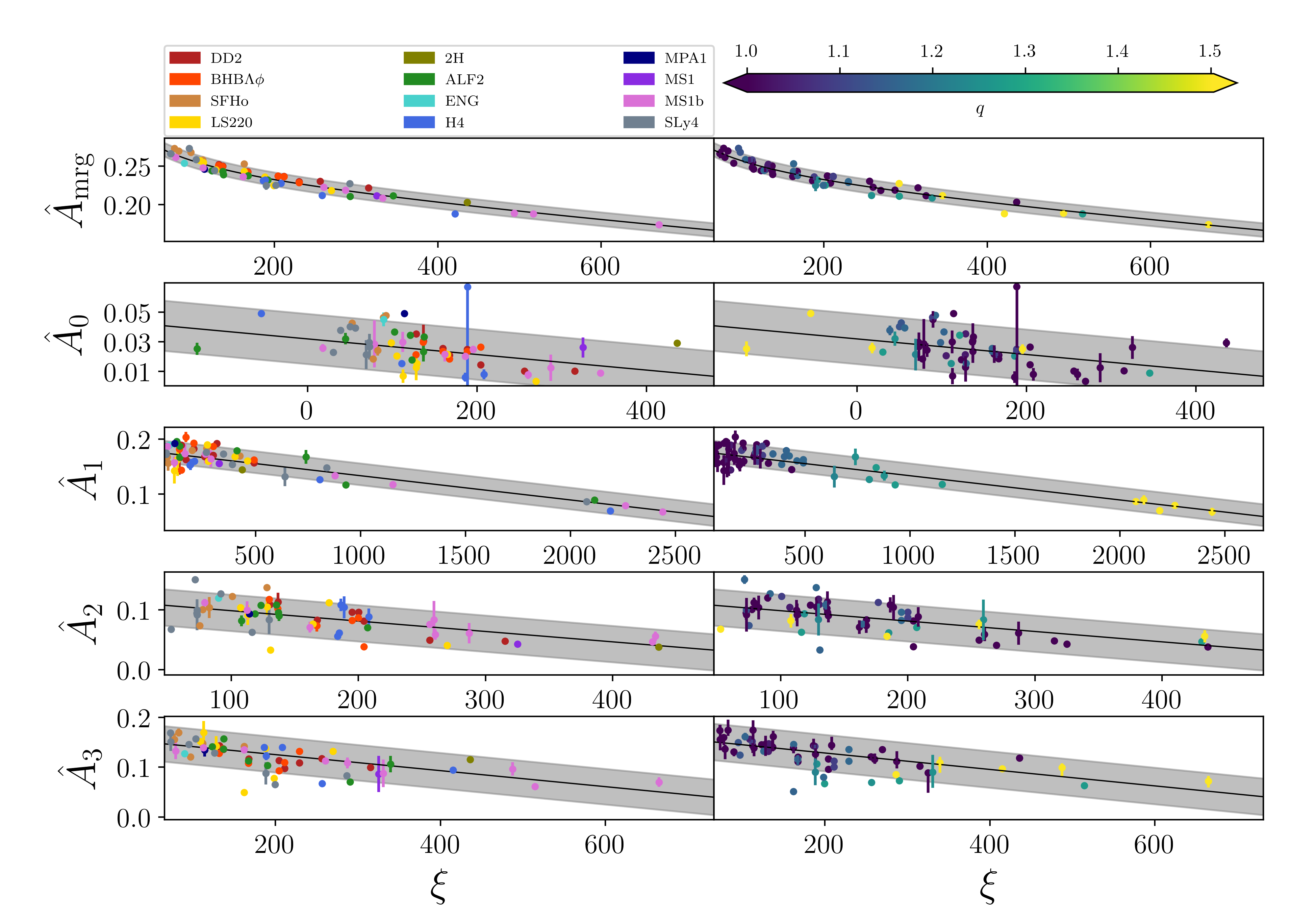}
  \includegraphics[width=.49\textwidth]{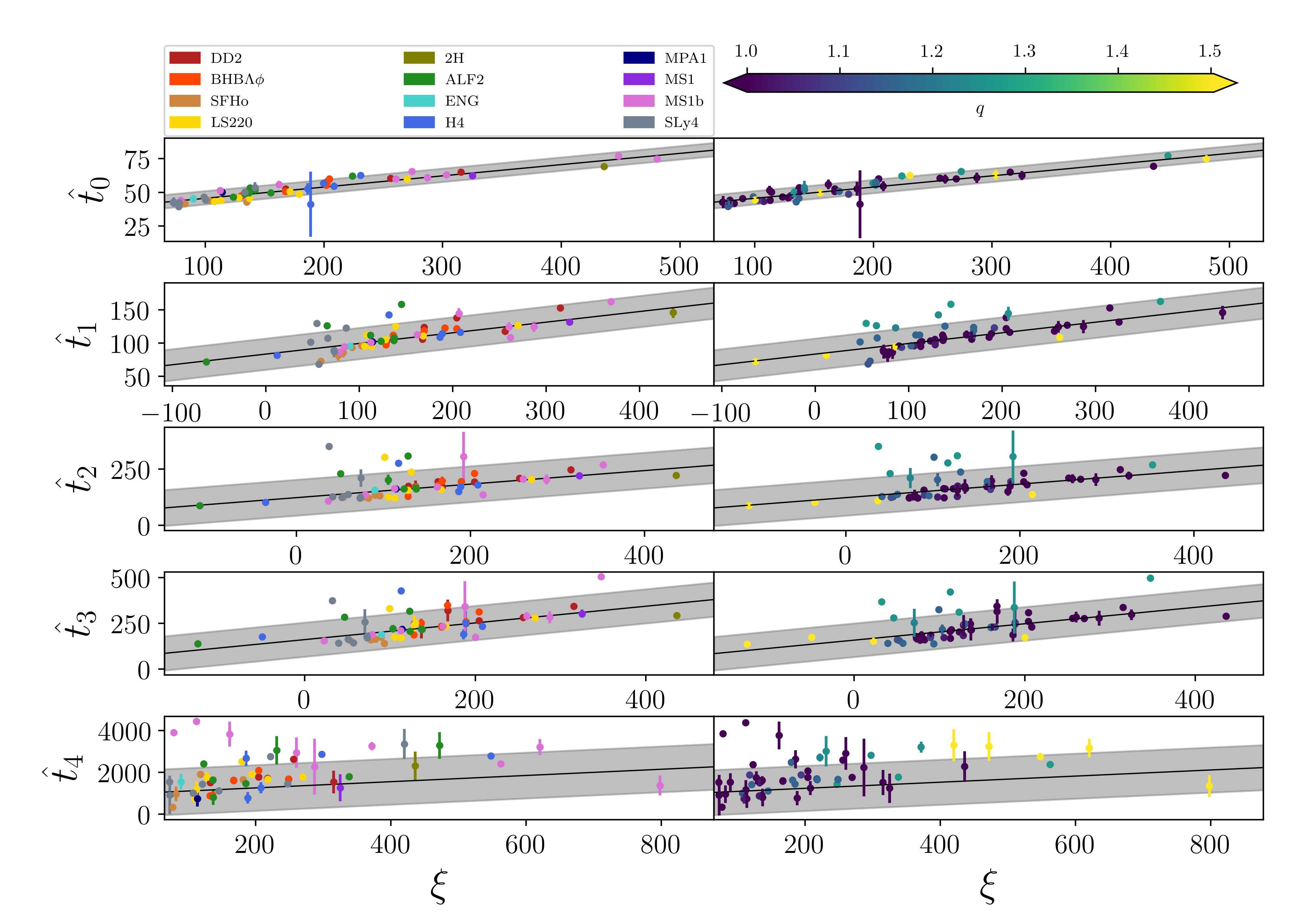}
    \caption{Characteristic amplitudes and times information from NR
      simulations. Markers represent the quantities 
      extracted from the NR data;
      the black lines are the fits with their 90\% credible regions.
      All upper panels show the same data; the colors on the left
      panels correspond to the EOS
      variation, on the right panel the mass ratio.
      Note that we impose a lower bound for $\hat A_0$ equal to zero
      for all those values of $\xi$ that lead to negative results in the fits.}
    \label{fig:amptimefit}
\end{figure*}

\begin{figure}[t]
  \centering 
    \includegraphics[width=.49\textwidth]{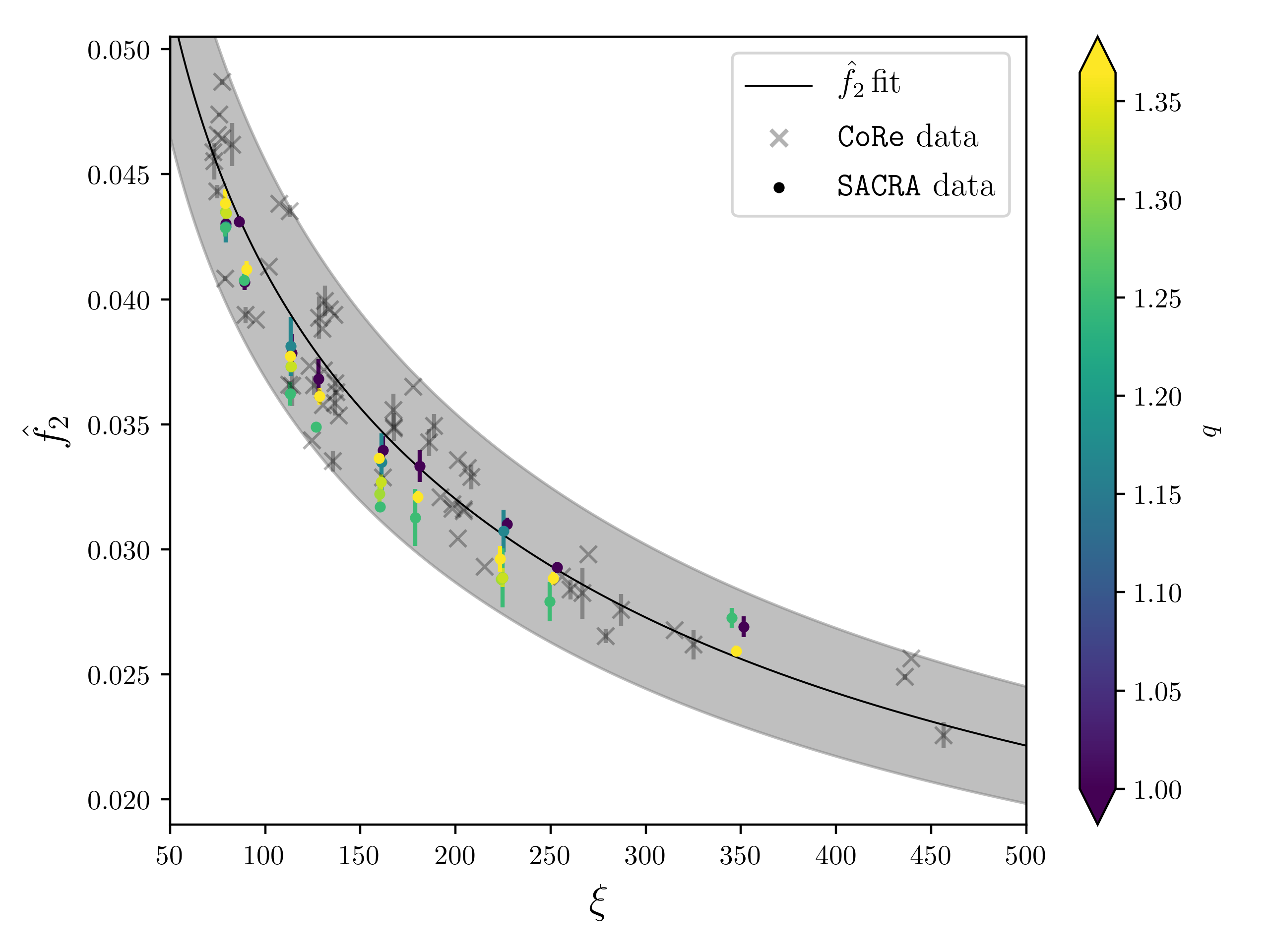}
    \caption{Postmerger frequencies $f_2$ from {\tt CoRe} database 
    (gray crosses) and from {\tt SACRA} catalog \cite{Kiuchi:2017pte, Kawaguchi:2018gvj} 
    (colored dots), averaged on different resolutions.
     The black solid line is the quasiuniversal relation for $\f_2$ extracted from {\tt CoRe} data with its 90\% credible region.}
    \label{fig:freqfit+sacra}
\end{figure}

We collect in this appendix various plots of quasinuniversal relations
for amplitudes and times. Fig.~\ref{fig:amptimefit} shows amplitudes 
and times fits extracted from NR data of {\tt CoRe} collaboration and 
implemented in \nrpm{} model. The robustness of those relations is
further demonstrated using the independent data from {\tt SACRA} code
\cite{Kiuchi:2019kzt} that were not used in this work.
To this purpose Fig.~\ref{fig:freqfit+sacra} shows 
a comparison between the $f_2$ extracted from the {\tt SACRA} catalog
\cite{Kiuchi:2019kzt} and the \texttt{CoRe} data and fits.

We give an euristic justification of the quasiuniveral relations
(employed here and elsewhere to summarize NR information) and of the
choice of the parametrization. The discussion follows from the
original argument given in \cite{Bernuzzi:2014owa}.

While the choice of the parameter in Eq.~\eqref{eq:xi} should be
primarily considered as an operative choice, it can be in part justified
based on perturbative arguments. In the effective-one-body (EOB) description
of the two-body dynamics or, equivalently in this case, in the
post-Newtonian formalism, the interbinary potential $A(u)$, where
$u=GM/(rc^2)$, is the main quantity which describes the binary dynamics.
The radial force governing the circular motion is given by
\be
\frac{dA}{dr} = -u^2\left(-2 + \hat{a}'_0(\nu,u)+
\hat{a}'_T(\kappa^A_{\ell},\nu,u)\right) \ ,
\ee
where, $\hat{a}_0$ and $\hat{a}_T$ are the point-mass and the tidal
corrections to the Newtonian term respectively (we neglect here spin interactions). The tidal
contribution is in general parametrized by the multipolar tidal
polarizability coefficients $\kappa^A_{\ell}$ of each NS \cite{Damour:2009wj}.
At leading order in $1/c^2$ the two terms above read
\be
\hat{a}_0(\nu,u)\propto \nu u^2 \ \ , \ \
\hat{a}_T(\kappa^A_{\ell},\nu,u)\propto - \kt2 u^3 \ \ .
\ee
Hence, finite mass-ratio and tidal effects are parametrized at leading
order by $\nu$ and $\kt2=\kappa^A_2+\kappa^B_2$. Note the two
contributions are associated with different powers in $u$ (different
post-Newtonian orders) and have opposite sign.

As noted in \cite{Bernuzzi:2014owa}, in the strong field regime (where
the expansion above is not accurate), and in particular close to the
EOB last stable orbit $u\sim 0.14$, the tidal term $\hat{a}_T$ can
become numerically comparable to $\hat{a}_0$ as $\kt2\sim\O(100)$. This reflects the
physical fact that the tidal term grows faster ($\sim 1/r^3$) at
small separations than the non-tidal one ($\sim 1/r^2$). 
Based on this picture, it is thus natural to interpret the NR data in
terms of $\kt2$ because the latter is the theoretically justified
parameter that encode the main effects of the EOS and masses on the
dynamics.

Interestingly, the $\kt2$ parameter approximately captures 
the collapse threshold and disk masses for nearly equal masses BNS 
\cite{Zappa:2017xba,Radice:2017lry}. On the one hand, this might
be intuitive since $\kt2$ contains information on the compactness of the
binary. On the other hand it is not necessarily expected, given that the
collapse is controlled by the maximum mass (pressure) supported by the
EOS at densities much higher than those of the individual NSs.
Thus, one should not expect the $\kt2$ parameter to completely or
accurately capture the strong field dynamics; for this reasons we
defined the NR relations as \textit{quasi}universal relations.
For example, to capture
the luminosity of binaries with mass-ratios significantly different
from unity, it is necessary to correct the leading-order
post-Newtonian coefficient by a function of $\nu$
\cite{Zappa:2017xba}. Similarly, in this paper we have introduced the
parameter $\xi$ in Eq.~\eqref{eq:xi} to better capture mass-ratio
effects. The logic behind Eq.~\eqref{eq:xi} is precisely to introduce
a term that can account for the strong-field effect of
$\hat{a}_0(\nu,u)$.
However, for the reasons above, the
$\xi$ parameter cannot properly describe quantities affected by significant
tidal disruption. An extreme case is for the example the disk mass in
BH-NS binaries \cite{Foucart:2018rjc,Zappa:2019ntl}. 


\section{Bayesian analysis with EOS inference} 
\label{app:EOS-inference-pc}

Constraints on the matter EOS can be extracted from the GW signal by
performing inference on a parameterized family of EOS
\cite{Lindblom:2010bb,Lackey:2014fwa,Abbott:2018exr,Carney:2018sdv}. 
Instead of sampling macroscopic EOS-related parameters,
we can directly sample the function $p(\rho)$ that defines the EOS.
Given this information, it is possible to infer the properties of each NS,
such as tidal parameters and radii.

This method can be applied also with the complete model proposed in this work,
as described in Sec.~\ref{sec:IMPM}. In this case, it is possible to use 
Eq.~\eqref{eq:Mthr} in combination with information from the EOS 
(instead of the tidal parameters and Eq.~\eqref{eq:k_thr}), in order 
to infer whether the remnant undergoes a prompt collapse to a BH. 
In particular, $\Mmax$ and the maximum NS compactness $C_{\rm max}$ would be calculated
from the inferred EOS while $k_\text{thr}$ can be estimated from the inferred $C_{\rm max}$ 
using NR fits~\cite{Bauswein:2013jpa,Agathos:2019sah}. 
This approach gives an alternative way to include prompt collapse in
complete waveform models based on \nrpm{} which we will further explore in
future work. 


\section{Robustness of NR postmerger waveforms} 
\label{app:long-conv}

\begin{figure}[t]
  \centering 
  \includegraphics[width=.49\textwidth]{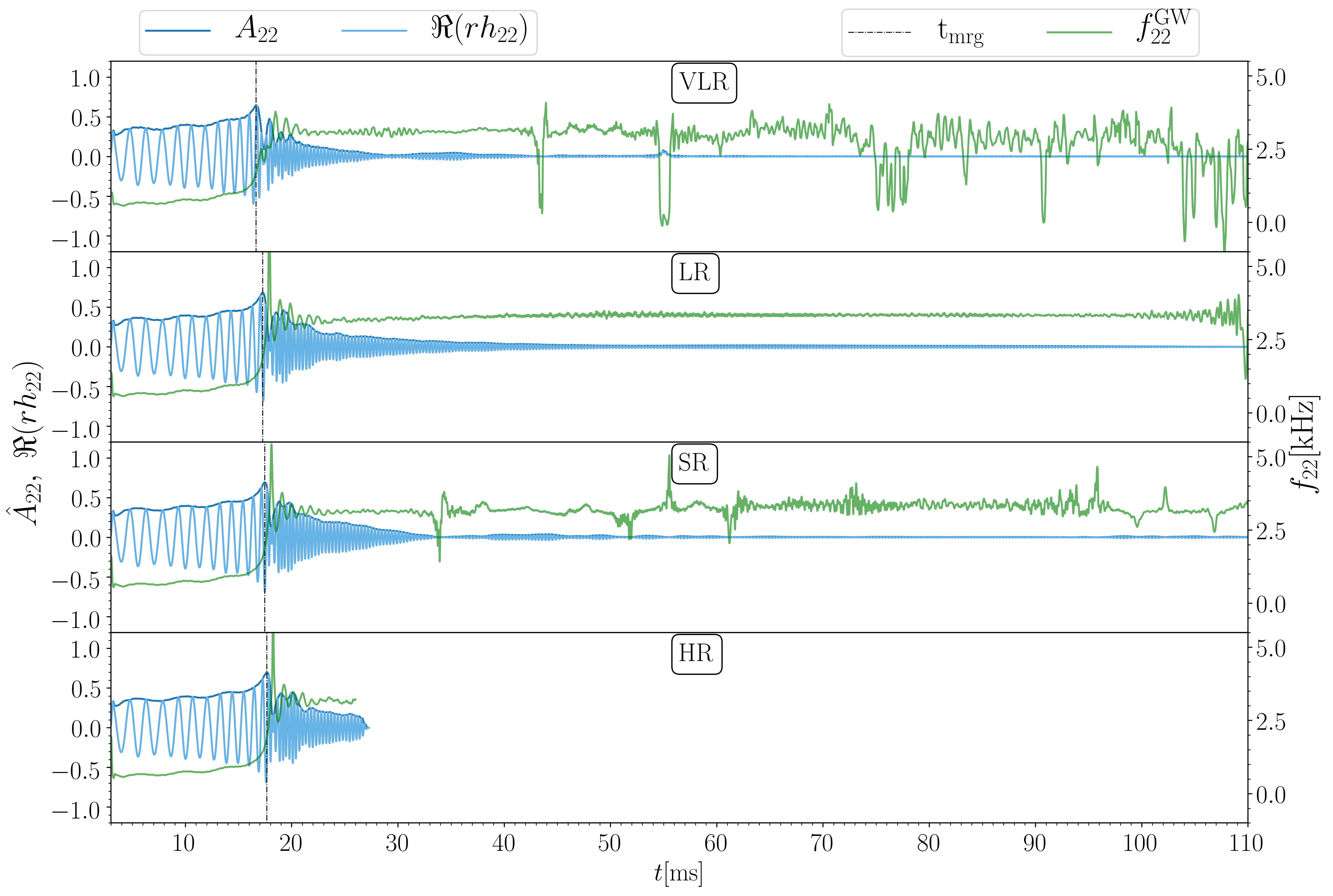}
    \caption{Dependence of NR waveform on the grid
      resolution for the simulation SLy4 $M= (1.30+1.30)\Mo$. 
      VLR, LR, SR, HR stand respectively for maximal resolutions $ h
      =[0.415,0.246,0.185,0.136]$~km in each direction.}
    \label{fig:NRconv}
\end{figure}

As discussed in the main text a main limitation in the construction of
accurate postmerger models is the quality of NR postmerger
waveforms. 
While the accuracy of inspiral-merger BNS waveforms has been studied in
some detail and clear waveform convergence can be shown using
high-order finite-differencing methods
\cite{Bernuzzi:2011aq,Bernuzzi:2012ci,Radice:2013hxh,Radice:2013xpa,Bernuzzi:2016pie},
the latter are less effective in postmerger simulations. Except for
notable cases \cite{Radice:2016gym,Radice:2016rys}, the robustness of
postmerger waveform with grid resolution has not been studied in
detail. We discuss here a resolution study of a long postmerger waveform.

Amongst the validation binaries, we simulated the evolution of the long-lived remnant employing a
microphysical EOS SLy4~\cite{daSilvaSchneider:2017jpg} starting from a
binary system of individual NS masses 
of $1.30~\Msun$ at different resolutions. 
These simulations span six orbits before merger and last for more than 100~ms after merger. 
Such integration times can be demanding in terms of
computational time but NR codes allow stable evolutions at rather
low grid resolution, e.g.~\cite{Andersson:2013mrx,DePietri:2018tpx,Ciolfi:2019fie,Nedora:2019jhl}. 
Evolutions are performed with the \texttt{WhiskyTHC} code
\cite{Radice:2012cu,Radice:2013hxh,Radice:2013xpa,Radice:2018xqa}
using a fifth-order monotonicity-preserving reconstruction within a
standard second order 
finite volume scheme~\cite{Radice:2013xpa}. Stars are covered with
resolutions of $h=[0.415,0.246,0.185,0.135]$~km in each direction,
respectively Very Low Resolution (VLR), 
Low Resolution (LR), Standard Resolution (SR), High Resolution (HR),
where SR is our standard for production runs 
\cite{Radice:2018pdn} (but note we
performed also several HR simulations in past work).
We use seven 2:1 refinement levels and Courant-Friderich-Lewy factor
of $0.075$ for the timestep. 

The $(2,2)$ waveforms from runs at different resolution are shown in
Fig.~\ref{fig:NRconv}. The waveform's amplitude has a non-monotonic
behavior with increasing resolution. For example, the extrema in the
time window $t\in(30,60)$~ms are similar for VLR and SR but different
from those of the LR data. The numerical high-frequency noise
affecting the frequency reduces in magnitude the higher the resolution
is, but it is mainly correlated to the 
amplitudes' minima. 
Hence, also the frequency noise is not converging with resolution at
the considered resolutions. We check the waveform phase convergence
and found that the phase has a monotonic behavior with the 
grid resolution only until few milliseconds after merger; the long-term
data are not in convergence regime at these resolutions.

Results at resolution VLR show the appearance of spurious frequencies
at $f<f_2$ around $40$~ms; the latter are not present at higher resolutions. 
These frequencies have been erroneously interpreted as physical
convective modes \cite{DePietri:2018tpx}, which are instead not developed on
these timescales even using a microphysical EOS. 
A careful inspection of the dynamics and multipolar waveform reveals instead physical spiral
modes with $m=1$ geometry 
\cite{Paschalidis:2015mla,East:2016zvv,Radice:2016gym,Lehner:2016wjg}.
The GW frequency of the mode is $f_1=f_2/2$ and could be added
to \nrpm{} model \cite{Lehner:2016wjg}, but it corresponds to a weak 
GW emission \cite{Radice:2016gym}.

We conclude that, to the best of the current knowledge, postmerger
waveforms on timescales of ${\sim}100$~ms are well described in terms of 
the frequencies and amplitudes modeled by \nrpm{}. The production of high-quality NR
postmerger waveforms is an urgent goal.



\bibliography{references,local}

\end{document}